\documentclass[a4paper, 12pt]{article}
\usepackage{amsmath}
\usepackage{amsfonts}
\usepackage{amssymb}
\usepackage{authblk}
\usepackage{booktabs}
\usepackage{cite}
\usepackage{float}
\usepackage{geometry}
\usepackage{graphicx}
\usepackage{mathrsfs}
\usepackage[version = 4]{mhchem}
\usepackage{siunitx}
\usepackage{tabularx}
\usepackage{chapterbib}		
\usepackage{listings}
\usepackage[colorlinks, linkcolor=blue, anchorcolor=black, citecolor=blue, urlcolor=blue, CJKbookmarks=False]{hyperref} 

\renewcommand{\vec}{\mathbf} 
\renewcommand{\Re}{\mathrm{Re}} 
\renewcommand{\Im}{\mathrm{Im}} 
\newcommand\abs[1]{\left|#1\right|}		
\newcommand\rbk[1]{\left(#1\right)}		
\newcommand\sbk[1]{\left[#1\right]}		
\newcommand\fbk[1]{\left\{#1\right\}}	
\newcommand\FT[1]{\mathcal{F}\left[#1\right]}		
\newcommand\iFT[1]{\mathcal{F}^{-1}\left[#1\right]}	
\newcommand\dd[2][]{\!\;\mathrm{d}^{#1}#2}       
\newcommand\pdv[2]{\frac{\partial #1}{\partial #2}}     

\title{4D-Explorer: A visual software for 4D-STEM data processing and image reconstruction}

\author[1]{Yiming Hu\thanks{These authors contributed equally: Yiming Hu and Si Gao.}}
\author[1,2]{Si Gao$^*$}
\author[3]{Xiaopeng Wu}
\author[1]{Xudong Pei}
\author[1]{Futao Huang}
\author[1]{Wei Mao}
\author[1]{Weiyang Zhang}
\author[3]{Aidan Horne}
\author[1]{Zhengbin Gu}
\author[1,3]{Peng Wang\thanks{Corresponding author: Peng Wang. Email address: \url{peng.wang.3@warwick.ac.uk}}}
\affil[1]{\it National Laboratory of Solid State Microstructures, Jiangsu Key Laboratory of Artificial Functional Materials, College of Engineering and Applied Sciences and Collaborative Innovation Center of Advanced Microstructures, Nanjing University, Nanjing 210093, China.}
\affil[2]{\it College of Materials Science and Engineering, Nanjing Tech University, Nanjing 210009, China.}
\affil[3]{\it Department of Physics, University of Warwick, Coventry CV4 7AL, UK.}

\date{June 12, 2023}

\begin{document}

\maketitle 
\begin{abstract}
	
	With the development of high-speed electron detectors, four-dimensional scanning transmission electron microscopy (4D-STEM) has emerged as a powerful tool for characterizing microstructures in material science and life science. However, the complexity of 4D-STEM data processing necessitates an intuitive graphical user interface software for researchers. In this regard, we have developed 4D-Explorer, an open-source, lightweight and extensible software for processing 4D-STEM data. It offers a visual and interactive workflow, including data preparation, calibration, image reconstruction and generating quantitative results. Furthermore, during calibration, our software includes a novel algorithm for rotational offset correction that uses a defocused 4D-STEM dataset and its axial bright field image, which has lower experimental requirements than conventional methods. We anticipate that 4D-Explorer will help researchers harness the capabilities of 4D-STEM technology.

\end{abstract}
\section{Introduction} \label{Section Introduction}

The state-of-the-art scanning transmission electron microscopes (STEM) have been widely used to image and characterize atom positions\cite{Batson2002Sub-Angstrom, Nellist2004Direct, Krivanek2003Towards}. Recently, owing to the development of high-speed pixelated electron detectors\cite{Plackett2013Merlin, Yang20154D-STEM, Tate2016High, Ciston20194D-Camera, Ercius20204D-Camera, Levin20204D-STEM, Haas2021High, Chatterjee2021Ultrafast, Jannis2022Event, Zambon2023KITE}, full diffraction images can be recorded in a short acquisition time. When the electron beam scans over the 2D real space (the sample plane), the fast detector records diffraction images in the 2D reciprocal space (the diffraction plane). These diffraction images make up the four-dimensional scanning transmission electron microscopy (4D-STEM) dataset\cite{Ophus2019Four}, which contains rich information that can be retrieved to reveal local interaction details between the electron beam and the sample after the acquisition process.

4D-STEM is a powerful technique that offers a range of imaging modes for various applications including virtual imaging\cite{Fundenberger2003Polycrystal, Watanabe2007Development, Caswell2009High, Gammer2015Diffraction, Lupini2015Ptychographic, Ophus2016Efficient, Hachtel2018Sub-Angstrom, Ahmed2020Visualization, Thronsen2022Studying}, positional averaged convergent beam electron diffraction (PACBED)\cite{LeBeau2010Position, Pollock2017Accuracy}, differential phase contrast (DPC)\cite{Rose1974Phase, Dekkers1974Differential, Hachtel2018Sub-Angstrom}, and electron ptychography\cite{Hoppe1969Diffraction, Rodenburg1992Theory,Nellist1995Resolution,Wang2017Electron,Gao2017Electron,Song2018Hollow,Jiang2018Electron,Song2019Atomic,Chen2020Mixed,Zhou2020Low,Chen2021Electron,Pei2023Cryogenic}. These techniques can be used to generate orientation mapping\cite{Panova2016Orientation, Sunde2018Evolution, Gallagher2019Nanoscale}, phase mapping\cite{Caswell2009High, Gammer2015Diffraction, Gallagher2019Nanoscale} and strain mapping\cite{Usuda2004Strain, Muller2012Strain, Ozdol2015Strain, Haas2017High, Pekin2017Optimizing, Pekin2018Insitu, Han2018Strain, Gammer2018Local}. One rapidly developing technique is center of mass (CoM) imaging which has the same physical origin as DPC using segmented detectors\cite{Muller2014Atomic, Gao2019Real}. CoM images are phase contrast images that are sensitive to light elements, and can also characterize electrostatic\cite{Muller2014Atomic} and magnetic fields of samples\cite{Kohno2022Real} with atomic-level precision. Meanwhile, the integrated CoM (iCoM) and differentiated CoM (dCoM) images can respectively provide information about the electrostatic potential and charge density of the sample\cite{Lazic2016Phase}.

Currently, there are various open-source software packages available for 4D-STEM data processing, such as py4DSTEM\cite{Savitzky2021py4dstem}, LiberTEM\cite{Clausen2020Libertem}, Pycroscopy\cite{Somnath2019UsidSmall}, Nion Swift\cite{Meyer2019Nion}, riCOM\cite{Yu2022Real} and PtychoSTEM\cite{Pennycook2019Efficient}. However, due to the complexity of 4D-STEM data processing, there are still some challenges in software development. First, the size of 4D-STEM datasets ranges from several GBs to TBs, which may exceed the typical memory capacity of personal laptops. Second, the high dimensionality of the data makes it difficult to directly view and interactively calibrate. Third, different research fields may require flexible and customizable data processing methods due to the diversity of experimental environments and sample types.

To address these challenges, we have developed an open-source software, 4D-Explorer. The software supports major operating systems (Windows, MacOS, Linux) and can be run on weaker hardware such as personal laptops with satisfactory performance. With 4D-Explorer, we achieved a complete and interactive 4D-STEM data processing workflow, resulting in various imaging modes. In addition, we demonstrated the impact of rotational offset correction and diffraction alignment on CoM imaging. We proposed a novel method to find the rotational offset angle using the axial bright field (Axial BF) image of a defocused 4D-STEM dataset. Furthermore, during image reconstruction, 4D-Explorer can calculate and display contrast transfer functions in real-time to assist users in optimizing reconstruction parameters and interpreting results. Our experimental results validated the effectiveness of the calibration and image reconstruction algorithms, resulting in quantitative imaging.

\section{Software Features} \label{Section Software Architectures and Features}

\begin{figure}[htbp]
	\includegraphics[width=\linewidth]{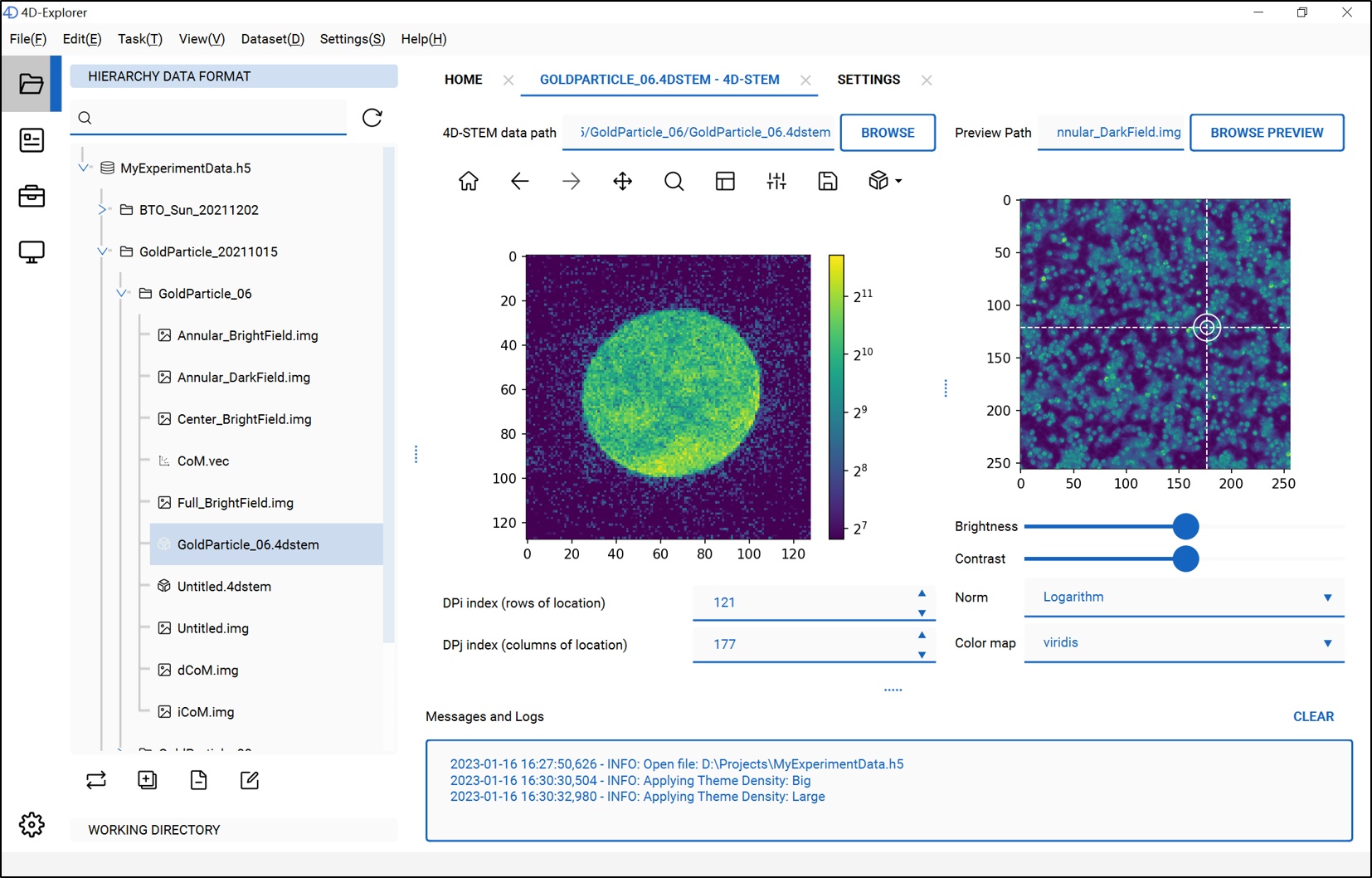}
	\centering
	\caption{A screenshot of the main window of 4D-Explorer. A tree view at the left displays datasets stored in an HDF5 file. The main window is showing the 4D-STEM dataset of gold nanoparticle-modified DNA sample. In the middle is a diffraction pattern, while the right half is the reconstructed annular dark field (ADF) image.}
	\label{Fig MainWindow}
\end{figure}

4D-Explorer, written in Python and utilizing open-source libraries from the scientific computation community, provides a visual framework for time-consuming data processing, including a file manager, a task manager, and several data viewers. Figure \ref{Fig MainWindow} is a screenshot of the main window of the software, and functions of the components are displayed in Supplementary Figure \ref{Fig Architecture}. Each component of the software addresses specific challenges in 4D-STEM data processing. The file manager and task manager handle the storage and computation of large datasets, while the data viewers provide flexible display options based on different purposes, processing methods and data dimensions.

\paragraph{File Manager} The file manager organizes and stores data, including lines, images, vector fields, and 4D-STEM datasets, all within a single HDF5\cite{Folk2011Overview} file, as shown in Supplementary Figure \ref{Fig File Manager}. It provides flexible data storage and high-performance input/output (IO), and loads data lazily to avoid memory overflow. The file manager handles different data types appropriately by distinguishing them through their respective file extensions, for example, ``.4dstem'' for 4D-STEM datasets, ``.vec'' for vector fields, and ``.img'' for 2D images. In addition, each data item in the file manager has its own set of attributes, which stores experimental parameters and other relevant information, as shown in Supplementary Figure \ref{Fig Attributes}. Users can view and edit the attributes as needed.

\paragraph{Task Manager} The task manager in 4D-Explorer schedules time-consuming computation processes in the background, preventing the GUI from being unresponsive. The task manager uses a queue-style scheduling algorithm to avoid conflicts that may arise when different tasks modify the same dataset. Users can view the progress, status, details, error messages, and history of these tasks (See Supplementary Figure \ref{Fig Control Panel} \textbf{III}). 

With both task manager and file manager, 4D-Explorer offers a 4D-STEM data processing with acceptable performance on handling big data using personal laptops. The time consumed by data loading and processing depends on the IO speed of the storage devices. For a \SI{4}{GB} 4D-STEM dataset stored in a hard disk drive (HDD), it costs \SI{48}{s} for image reconstruction and \SI{127}{s} for calibration, which depend on the reading speed and writing speed of the disk repectively. However, if the dataset is stored in a solid state drive (SSD), the time is reduced to \SI{17}{s} and \SI{21}{s} for reconstruction and calibration repectively. As the disk input/output (IO) becomes the limiting factor, the CPU utilization remains below $10\%$, and the memory usage is kept below \SI{130}{MB}. Generally, the time consumption scales linearly with the size of the dataset.

\paragraph{Viewer} A viewer in 4D-Explorer is a page designed to display different types of data according to their specific display requirements and purposes. Each viewer embeds Matplotlib windows\cite{Hunter2007Matplotlib} for interactive, multiplatform, and high-quality visualization. There are currently three types of base viewers: image viewer, vector field viewer, and 4D-STEM viewer. The image viewer allows us to adjust display effects such as brightness, contrast, and interpolation algorithms, as well as pan and zoom-in on displaying regions and view its histogram (Supplementary Figure \ref{Fig Image Viewer}). The vector field viewer displays quiver plots with a background image that visually demonstrates the distribution of electric fields at the atomic level along with their potential or charge density (Supplementary Figure \ref{Fig Vector Field Viewer}). Meanwhile, the 4D-STEM viewer can simultaneously display diffraction patterns and a preview image (Figure \ref{Fig MainWindow}). By interactively dragging the cursor on the preview image, we can explore different diffraction patterns. This preview image's correspondence with the diffraction patterns facilitates an intuitive understanding of the 4D-STEM dataset and potential problem diagnosis.

\paragraph{Extensibility} The architecture and the python open-source scientific computing ecosystem provide the functional extensibility of 4D-Explorer. Users are able to handle the HDF5 dataset directly by many other open-source tools when they need advanced processing. Among these tools, LiberTEM\cite{Clausen2020Libertem} provides graphic interfaces based on Jupyter notebook and parallel imaging algorithms following the Map-Reduce model, but currently lacks visual calibration functions. Therefore, we can first use 4D-Explorer to calibrate 4D-STEM datasets and then use LiberTEM for parallel reconstruction, especially in the cases where the process becomes computationally intensive. Other software packages such as pyXem\cite{Duncan2022Pyxem} and Pycroscopy\cite{Somnath2019UsidSmall} contain comprehensive calibration and reconstruction algorithms, but they are limited in the aspect of graphical interface. As a complementation, 4D-Explorer provides convenient methods for primary analysis that help users to quickly detect imaging defects and select customized calibration algorithms. Moreover, The GUI and file manager also provide an opportunity for users to implement customized algorithm and method development without concern for data display and IO issues.

Overall, 4D-Explorer is designed with both usability and flexibility in mind. With the help of the file manager, task manager, and various viewers, this tool can significantly lower the barriers for researchers from different fields who wish to explore the world of 4D-STEM techniques and quickly obtain results.

\section{Workflow}\label{Section Workflow}

\begin{figure}[htbp]
	\includegraphics[width=\linewidth]{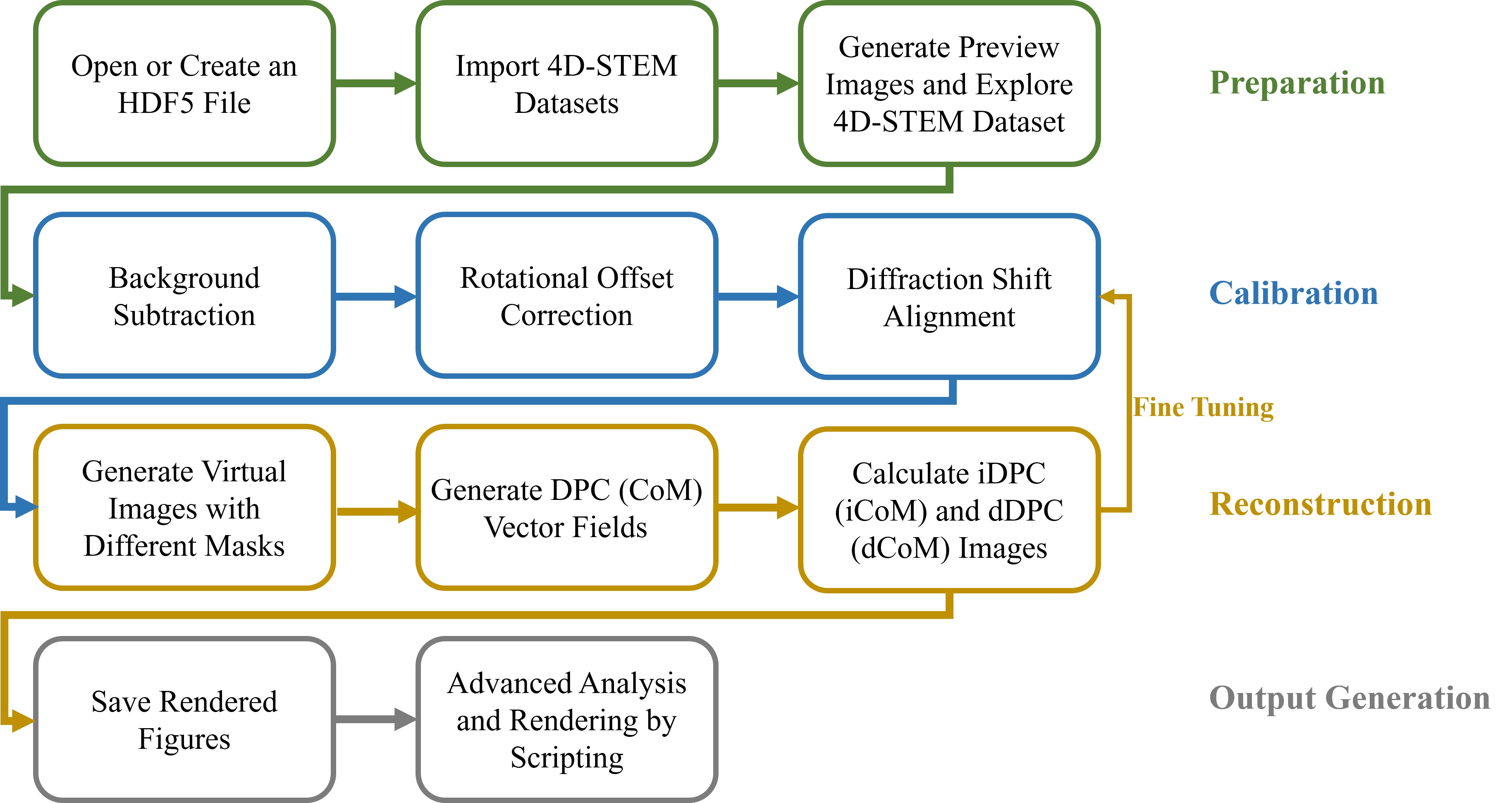}
	\centering
	\caption{Workflow for 4D-STEM datasets analysis in 4D-Explorer. First in the preparation step, we load raw 4D-STEM dataset into the file manager. Then, we calibrate the dataset for physically meaningful and accurate images. Next, reconstruction algorithms are used to compute images. These images can also be used as basis for fine tuning. Finally results are rendered by appropriate viewers. They can then be displayed and saved for advanced analysis.}
	
	\label{Fig Workflow}
\end{figure}

The workflow of 4D-STEM dataset processing typically involves four major steps: preparation, calibration, reconstruction, and output generation, as illustrated in Figure \ref{Fig Workflow}. In this section, we demonstrate the calibration step using a gold nanoparticle-modified DNA sample, and high-resolution image reconstruction step using a monolayer \ce{MoS_2} sample. The 4D-STEM experiments are performed using an aberration-corrected FEI Titan$^3$ Cubed 60-300 STEM equipped with an electron microscopy pixel arrayed detector (EMPAD). The experimental parameters used for the acquisition are listed in Table \ref{Table Optical configuration}. 

\subsection{Preparation} 

To begin with the preparation step, we first create a new HDF5 file. Then, the raw 4D-STEM datasets can be imported into the HDF5 file via the file manager. Currently, the software can automatically import RAW files from EMPAD\cite{Tate2016High} and MIB files from Merlin Medipix3\cite{Plackett2013Merlin} (Supplementary Figure \ref{Fig Importer}). The experimental parameters of the 4D-STEM dataset, such as accelerating voltage $U$, convergent semi-angle $\alpha$, camera length (CL), and scanning step size, can be automatically parsed and stored as attributes of the dataset (Supplementary Figure \ref{Fig Attributes}). These experimental parameters can also be manually edited and will be useful in later steps of the workflow.

\subsection{Calibration}

Calibration is an essential step to acquire accurate and meaningful 4D-STEM data. It is not uncommon for the instrument to be not well calibrated, which can lead to inaccuracies in the calculated parameters. To obtain quantified 4D-STEM results, neither of the two calibration procedures, rotational offset correction and diffraction alignment, can be neglected. In addition to these two procedures, 4D-Explorer also provides user-friendly interfaces for other calibration options, such as background subtraction (Supplementary Figure \ref{Fig Background Subtraction Viewer}). Further development and implementation of calibration algorithms and methods will continue to be explored to improve the accuracy and reliability of 4D-STEM results.

\paragraph{Rotational Offset Correction} 

\begin{figure}[htbp]
	\includegraphics[width=\linewidth]{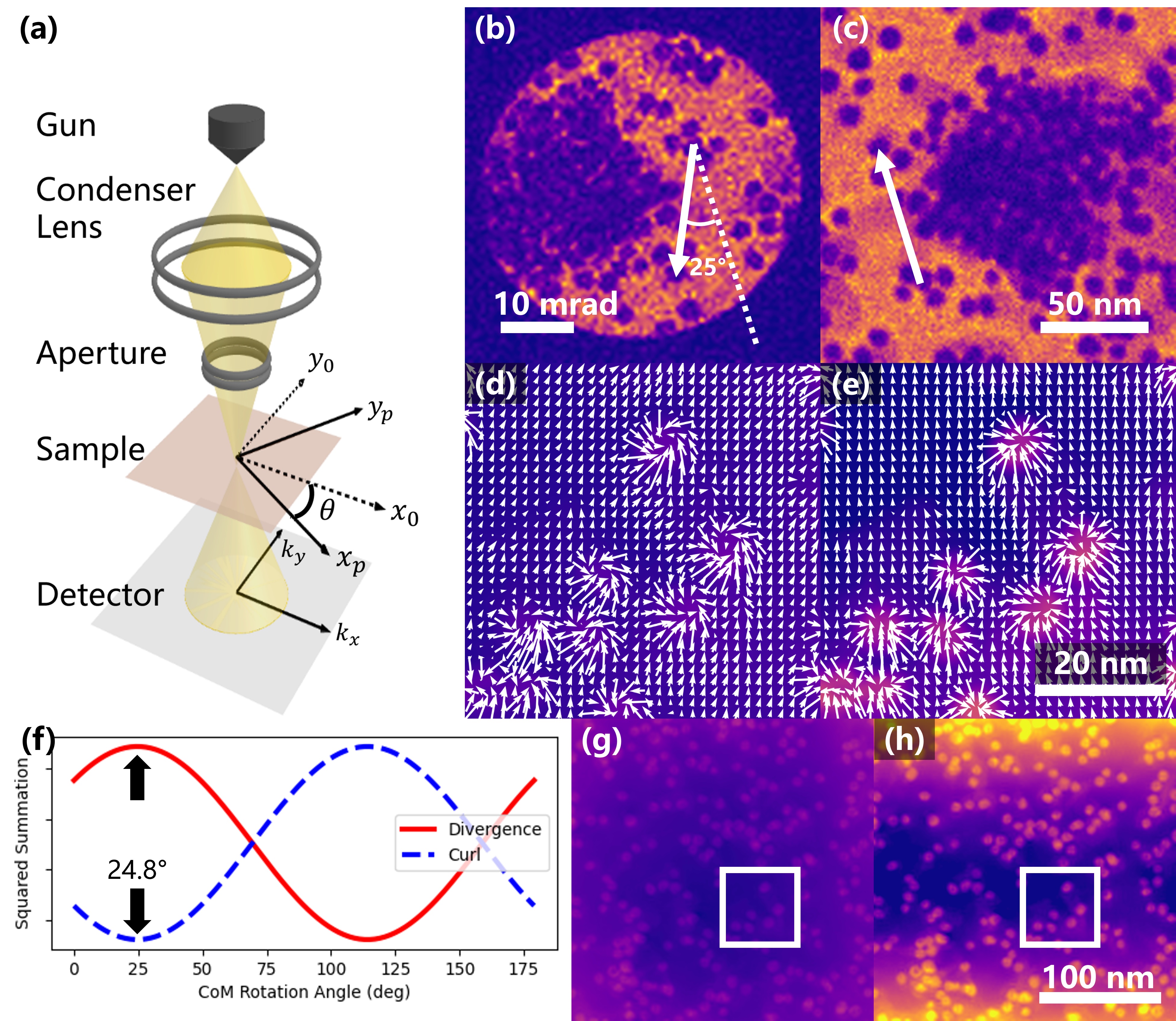}
	\centering
	\caption{Rotational offset correction. (\textbf{a}) Schema of rotational offset during the 4D-STEM experiment. The probe is scanning along the $x_0$ and $y_0$ axes, which are not parallel to respective axes of the pixel arrays of the detector ($k_x$ and $k_y$). (\textbf{b},\textbf{c}) An underfocused 4D-STEM dataset of gold nanoparticles, including (\textbf{b}) diffraction pattern and (\textbf{c}) virtual Axial BF image. (\textbf{d},\textbf{e},\textbf{g},\textbf{h}) Focused 4D-STEM dataset of gold nanoparticles before and after rotational offset correction, including (\textbf{d}) CoM before correction, (\textbf{e}) CoM after correction, (\textbf{g}) iCoM before correction and (\textbf{h}) iCoM after correction. (\textbf{f}) divergence and curl curve of the CoM field with respect to the rotation angle.} 
	\label{Fig Rotational Offset}
\end{figure}

The rotational offset in electron microscopes, which refers to the rotation between the scanning coordinate and the detector coordinate, arises from three sources shown in Supplementary Figure \ref{Fig Rotational Offset Source}. Firstly, the helical trajectory of electrons caused by magnetic lenses induces a rotation in the electron beam, which in turn causes a rotation in the formed image\cite{Williams2009Transmission}. Secondly, different choices of coordinate basis during computing, such as using $x-y$ coordinates versus $i-j$ indices, can lead to an additional rotation. Lastly, manually set scanning rotations in the experiment also contribute to the overall rotational offset. These three angles from the sources can be summarized as $\theta$, as shown in Figure \ref{Fig Rotational Offset}\textbf{a}. Although the rotational offset does not affect virtual imaging, its correction is essential for modes that are determined by the orientations of diffraction patterns, such as strain mapping, orientation mapping, DPC and CoM. To measure the rotational offset, 4D-Explorer provides following two methods. 

The first method to measure the rotational offset angle is to compare the sample features between the real-space scanning image and the shadow image, which contains the sample shape information in the diffraction patterns. In a convention approach of this method\cite{Savitzky2021py4dstem}, the real-space scanning image can be an in-focus STEM image, while the shadow image is acquired from a defocused diffraction image. Here we propose a novel approach that replaces the in-focus STEM image with the Axial BF image reconstructed from a defocused 4D-STEM dataset. The Axial BF image is a good approximation of the conventional TEM (CTEM) image due to the reciprocity theory\cite{Pogany1968Reciprocity}, and thus it has a large depth of fields. Therefore, it can preserve the sample features when the defocus is as large as about \SI{2}{\micro\meter} in our experiment. Compared to the conventional approach, our approach only requires one single defocused 4D-STEM dataset, which simplifies the experimental operation and avoids the challenge of drifting that may arise during changes in defocus values. However, it should be noted that the shadow image of underfocus and overfocus datasets may be rotated \SI{180}{\degree} with respect to each other. Moreover, the sample features in the shadow images may be distorted by a large aberration, which could lead to an error in the measurement of $\theta$.

The second method is to compute the curl of the CoM (Supplementary Figure \ref{Fig Rotational Offset Viewer}). This method can be used when a defocused pairing dataset for calibration is not accessible, as it only requires the experimental 4D-STEM dataset itself. When the thin phase object approximation holds, we can obtain the CoM field distribution of the sample. The CoM field should be irrotational, since it is proportional to the gradient of the projected electric potential\cite{Lazic2016Phase}:
\begin{equation}
    \vec{I}_{\text{CoM}}(\vec{r}_p) = \frac{1}{2\pi}\left[\left|\psi_{\text{in}}(\vec{r})\right|^2\star\nabla\phi(\vec{r})\right](\vec{r}_p)
\end{equation}
where $\psi_{\text{in}}$ is the wave function of electron probe, $\phi$ is the modulating phase-change function of the sample, $\vec{r}$ is a pseudo-variable denoting locations of the sample, and $\vec{r}_p$ denotes the positions on the scanning plane. The $\star$ operator denotes cross-correlation\footnote{Cross correlation is defined as for $f(\vec{r})$ and $g(\vec{r})$, $\left[f(\vec{r})\star g(\vec{r})\right](\vec{r}') = \int_{\mathbb{R}^2} f^*(\vec{r})g(\vec{r}' + \vec{r})d^2\vec{r}$.} here. Note that $\phi$ is actually proportional to the projected electric potential of the sample. Then the angle can be determined by rotating each CoM vector and minimizing the sum of the squared curl:
\begin{equation}
    \theta = \arg\min_{\theta}\int_{\mathbb{R}^2}|\nabla_p\times\vec{I}_{\text{CoM}}(\vec{r}_p;\theta)|^2 d^2\vec{r}_p
\end{equation}
where $\nabla_p\times$ is the curl operator in the offset coordinate. The theory of the CoM-based method can be found in Supplementary Information \ref{Appendix Rotaional Calibration}. A similar method is also mentioned in \cite{Savitzky2021py4dstem}, and it has the same physical origin as the J-Matrix method introduced by \cite{Ning2022Accurate}.

In Figure \ref{Fig Rotational Offset}, we present the process of rotational offset correction using the two methods. In the first method, the axial BF in Figure \ref{Fig Rotational Offset}\textbf{c} is sharp and clear enough for matching features to the diffraction pattern in Figure \ref{Fig Rotational Offset}\textbf{b} and resulting in \SI{25}{\degree} as the rotational offset angle, while the conventional ABF and ADF images appear blurred as shown in Supplementary Figure \ref{Fig Unblurred Axial BF}. We note that for underfocused datasets, the diffraction image will be rotated an additional \SI{180}{\degree}, as evidenced in Figure \ref{Fig Rotational Offset}\textbf{b}, where the white arrow is rotated \SI{205}{\degree} compared to the arrow in Figure \ref{Fig Rotational Offset}\textbf{c}. In the second method, on the other hand, we use the curl of the CoM field to determine the optimal rotational offset angle. The curl of CoM reaches minimum at an angle of \SI{24.8}{\degree} in Figure \ref{Fig Rotational Offset}\textbf{f}, coinciding with the result from the first method. The iCoM images before and after the correction are shown in Figure \ref{Fig Rotational Offset}\textbf{g} and Figure \ref{Fig Rotational Offset}\textbf{h}, while Figure \ref{Fig Rotational Offset}\textbf{d} and Figure \ref{Fig Rotational Offset}\textbf{e} correspond to the CoM fields in the region of the white square in Figure \ref{Fig Rotational Offset}\textbf{g} and Figure \ref{Fig Rotational Offset}\textbf{h}, respectively. After the correction, the vector arrows of CoM in Figure \ref{Fig Rotational Offset}\textbf{e} point towards the centers of gold nanoparticles, consistent with our expectation. However, there are still artifacts in the iCoM image Figure \ref{Fig Rotational Offset}\textbf{h} after the correction, which will be eliminated by further diffraction alignment.

\paragraph{Diffraction Alignment} 

Diffraction alignment plays an important role in the accuracy of CoM images by moving the optical axis to the center of the diffraction images. The CoM of each diffraction pattern can be offset from the digital center due to various reasons, including the sample electric field scattering the electron beam, optical axis misalignment, and scanning induced diffraction wobble\cite{Craven1981Design}. To obtain a precise electric field distribution of the sample, it is necessary to reduce the impact of the latter two factors. Notably, the problem of diffraction wobble becomes critical when the scanning field of view exceeds tens of nanometers and may result in artifacts in the CoM results\cite{Savitzky2021py4dstem}.

For the misalignment of the optical axis, 4D-Explorer provides a visualization window to manually adjust the position of the bright field diffraction disk, so that it can be shifted to the center of the image, as shown in Supplementary Figure \ref{Fig Diffraction Alignment Viewer}. Subsequently, the measured shift value can be applied to all diffraction patterns in the 4D-STEM dataset to accomplish the diffraction alignment.

Diffraction wobble, however, results in varying displacements of the bright field diffraction disks relative to the center of each image, and thus each diffraction image requires individual measurements. In 4D-Explorer, two methods are available for measuring the displacements. The first method is to acquire a reference 4D-STEM dataset without the sample under the same experimental conditions, and calculating its CoM vector distribution. This CoM vector distribution is then applied as the translation vectors to align the experimental dataset with the sample.

\begin{figure}[htbp]
	\includegraphics[width=0.8\linewidth]{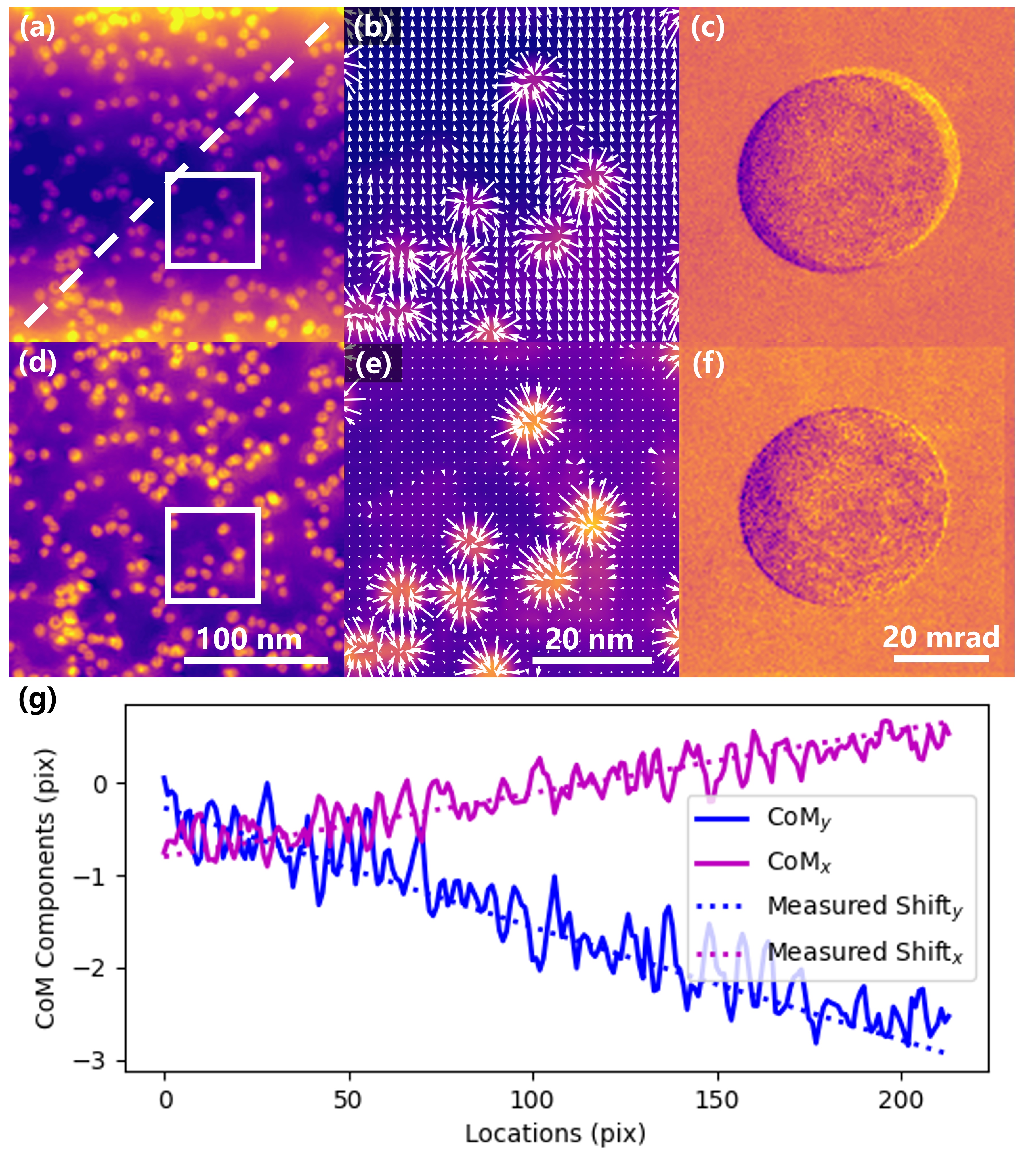}
	\centering
	\caption{Diffraction alignment. (\textbf{a}-\textbf{c}) before alignment, (\textbf{a}) iCoM image. (\textbf{b}) the zoomed-in CoM distribution with iCoM as background of (\textbf{a}). (\textbf{c}) difference between diffraction patterns at the both ends of the dashed line in (\textbf{a}). (\textbf{d}-\textbf{f}) after alignment, corresponding images to (\textbf{a}-\textbf{c}). (\textbf{g}) line profile of CoM components and measured shift curve with respect to the dashed line in (\textbf{a}).}
	\label{Fig Alignment}
\end{figure}

The second method to address the diffraction wobble issue is fitting, when no reference dataset is recorded in the experiment. We can select scanning points where there are no sample structures, and locate the center of the bright field diffraction disk. Then, we fit a linear or polynomial function through the displacement of the selected points to determine the translation vectors. In Figure \ref{Fig Alignment}, we present the result of CoM reconstruction before and after the alignment process. The input dataset used for this process is the outcome of the rotational offset correction we mentioned earlier. Prior to alignment, the presence of diffraction wobble in the dataset result in large-scale contrast artifacts in the iCoM image Figure \ref{Fig Alignment}\textbf{a}, which also contribute to the background vectors in Figure \ref{Fig Alignment}\textbf{b}. Additionally, in Figure \ref{Fig Alignment}\textbf{c}, we can observe that the locations of the bright field diffraction disk at the ends of the dashed line in Figure \ref{Fig Alignment}\textbf{a} do not coincide. In Figure \ref{Fig Alignment}\textbf{g}, we draw the distributions of the $x$ and $y$ components of CoM vectors and their corresponding fit functions along this dashed line. After applying the translation to each diffraction image, the artifacts in the iCoM image are eliminated in Figure \ref{Fig Alignment}\textbf{d}, the background vectors in Figure \ref{Fig Alignment}\textbf{e} are subtracted, and the diffraction disks are aligned and registered in Figure \ref{Fig Alignment}\textbf{f}.

In summary, the calibration process employed in 4D-Explorer involves multiple steps and methods, such as rotational offset correction and diffraction alignment, to ensure high-quality 4D-STEM imaging, particularly CoM imaging. Throughout the calibration process, the CoM fields of the gold nanoparticles modified DNA evolve from Figure \ref{Fig Rotational Offset}\textbf{d} to Figure  \ref{Fig Rotational Offset}\textbf{e}, and then from Figure \ref{Fig Alignment}\textbf{b} to Figure \ref{Fig Alignment}\textbf{e}, ultimately resulting in a clear electrostatic field image of the sample. In parallel, the iCoM images of the sample also undergo a transformation from Figure \ref{Fig Rotational Offset}\textbf{g} to Figure \ref{Fig Rotational Offset}\textbf{h}, and then from Figure \ref{Fig Alignment}\textbf{a} to Figure \ref{Fig Alignment}\textbf{d}, resulting in a phase contrast image of both gold nanoparticles and DNA molecules.

\subsection{Reconstruction} 

With the rich information provided by 4D-STEM datasets, specific reconstruction methods are required to generate and interpret images. These methods are necessary to extract the full potential of the data and provide a comprehensive understanding of the sample. The 4D-Explorer software currently offers two reconstruction modes: virtual imaging and center of mass. However, more algorithms are planned for future releases to further enhance the capabilities of 4D-STEM imaging. 

\paragraph{Virtual Imaging}

\begin{figure}[htbp]
	\includegraphics[width=\linewidth]{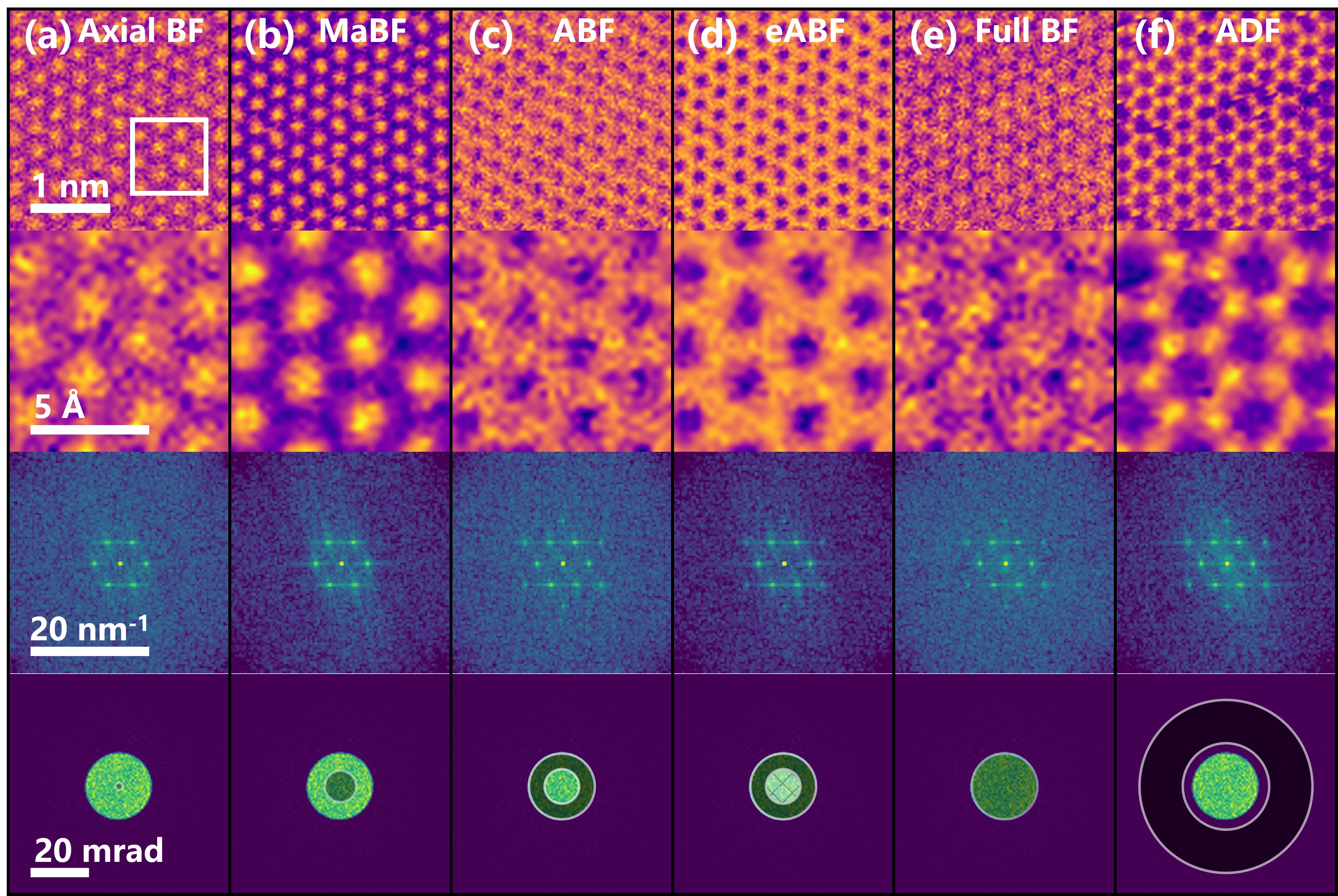}
	\centering
	\caption{Virtual Imaging. Reconstructions using (\textbf{a}) axial bright field (Axial BF), (\textbf{b}) middle angle bright field (MaBF), (\textbf{c}) annular bright field (ABF), (\textbf{d}) enhanced annular bright field (eABF), (\textbf{e}) full bright field (Full BF) and (\textbf{f}) annular dark field (ADF). The four rows are reconstructed images, the zoomed-in images of the white square region in (\textbf{a}), power spectrums and diffraction patterns with virtual detectors (ROIs), respectively.}
	\label{Fig Virtual Images}
\end{figure}

To generate virtual images, different regions of interest (ROIs) can be selected as virtual detectors on the diffraction patterns. The signals from the diffraction patterns within each ROI are then integrated and mapped onto the scanning coordinate system to form a virtual imaging result. The properties of virtual imaging modes are consistent with those of conventional STEM images when the virtual detector and conventional STEM detector have the same collection angle. In Figure \ref{Fig Virtual Images}, there are some common virtual imaging modes include bright field (BF), annular bright field (ABF) and annular dark field (ADF) of \ce{MoS_2} sample. In addition to these modes, Axial BF, middle angle bright field (MaBF)\cite{Findlay2014Enhanced} and enhanced annular bright field (eABF)\cite{Findlay2014Enhanced} can also be calculated. These modes provide additional flexibility in generating virtual images and can be used to extract specific information from the 4D-STEM dataset. 

\paragraph{Center of Mass}

\begin{figure}[htbp]
	\includegraphics[width=0.67\linewidth]{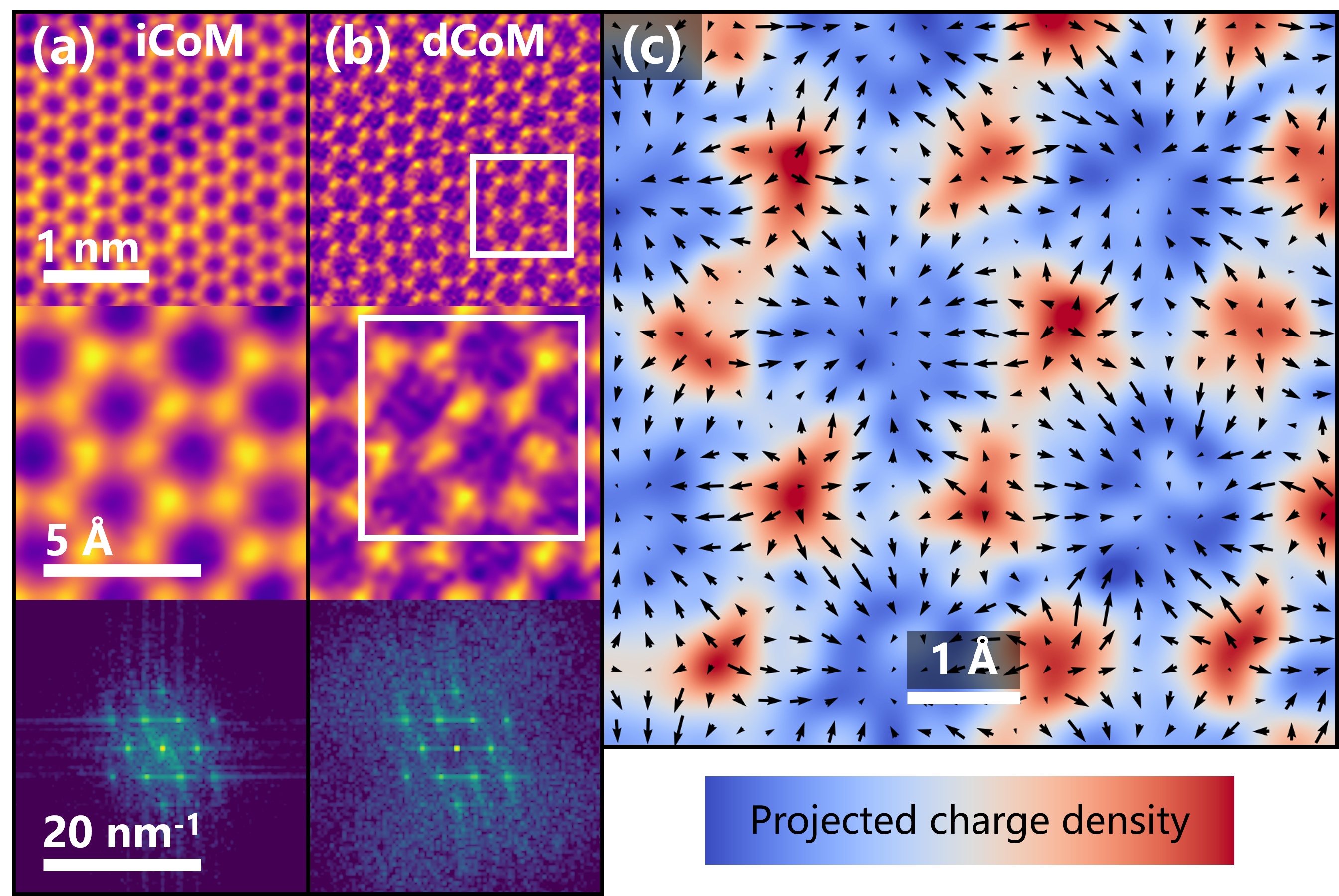}
	\centering 
	\caption{CoM reconstruction. (\textbf{a}) Differentiated CoM (dCoM). (\textbf{b}) Integrated CoM (iCoM). The three rows are reconstructed images, the zoomed-in images of the white square region in (\textbf{a}), and power spectrums. (\textbf{c}) CoM vector fields with projected charge density as the background image.}
	\label{Fig CoM}
\end{figure}

The distribution of the CoM vector field is obtained by computing the center of mass of each diffraction pattern, which reveals the momentum transfer of the electron beam caused by internal electric fields in the sample. For thin, non-magnetic samples, the CoM vector field is proportional to the projected local electric field. As the potential of the CoM field, integrated CoM (iCoM) is proportional to the projected electric potential. On the other hand, as the divergence of CoM vector field, differentiated CoM (dCoM) is proportional to the projected charge density according to Maxwell's Theorem. Figure \ref{Fig CoM} shows atomic-resolution CoM reconstructions of the \ce{MoS_2} sample. Figure \ref{Fig CoM}\textbf{a} and Figure \ref{Fig CoM}\textbf{b} show dCoM and iCoM reconstructions, respectively, while Figure \ref{Fig CoM}\textbf{c} displays the distribution of projected electric field with projected charge density as background. 

\paragraph{Contrast Transfer Functions} 

\begin{figure}[htbp]
	\includegraphics[width=\linewidth]{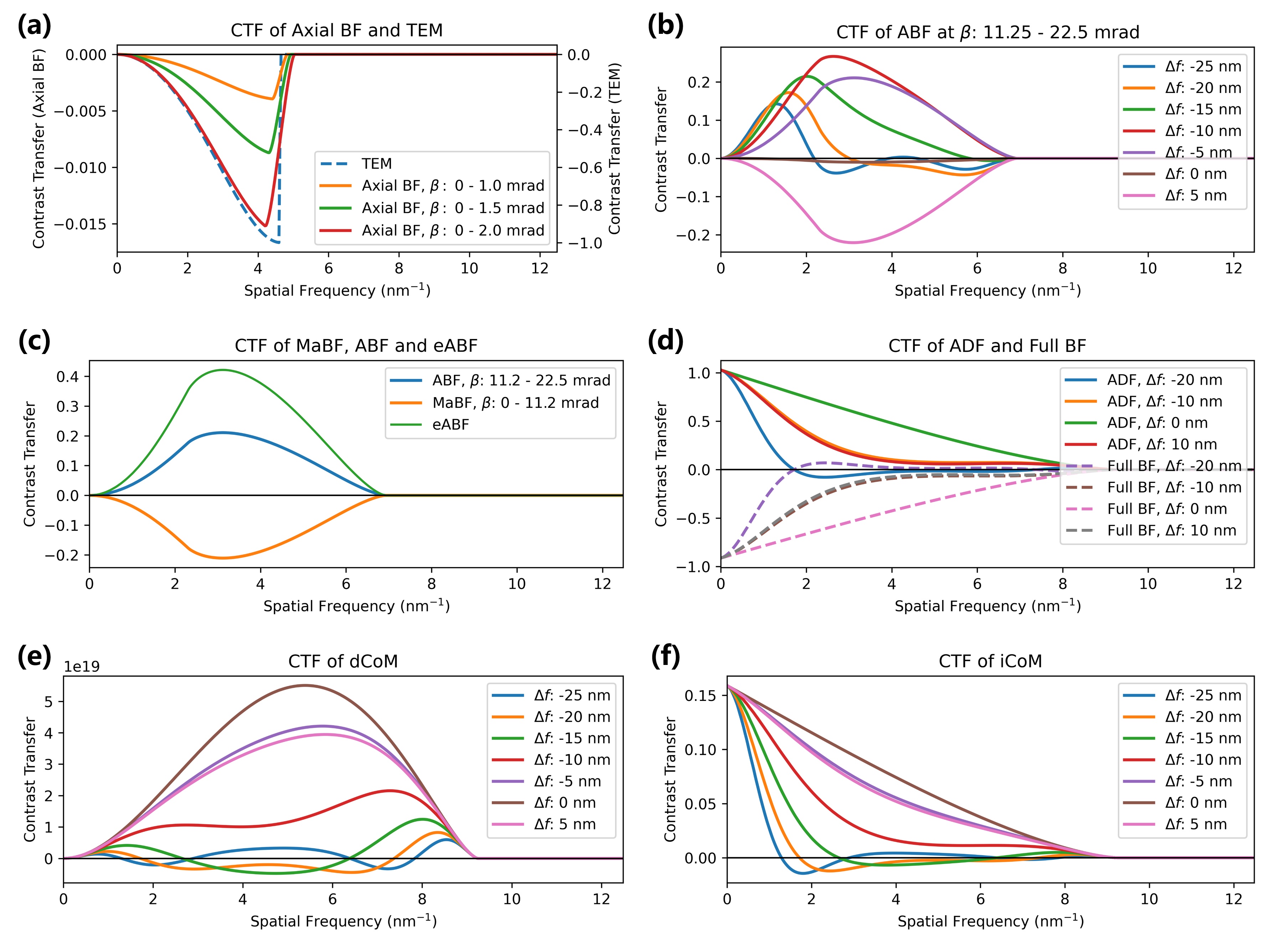}
	\centering
	\caption{Contrast Transfer Functions of Virtual Imaging. The convergent semi-angle $\alpha = \SI{22.5}{mrad}$, corresponding to spatial frequency $\SI{4.62}{nm^{-1}}$. In all the calculations, we added an spherical aberration $C_S = \SI{1}{\micro\meter}$. (\textbf{a}) Axial BF, LaBF, MaBF and TEM (in reciprocity) with defocus $\Delta f = -\SI{5}{nm}$. (\textbf{b}) ABF using collection semi-angle $\beta$ from $\SI{11.25}{mrad}$ to $\SI{22.5}{mrad}$. (\textbf{c}) MaBF, ABF and eABF with defocus $\Delta f = -\SI{5}{nm}$. (\textbf{d}) Full BF and ADF. The CTFs of ADF are calculated using collection semi-angle $\beta$ from $\SI{27}{mrad}$ to $\SI{67.5}{mrad}$. Note that the CTF of Full BF and ADF are calculated respect to the object $1-\cos\phi$, which approximates to $\frac{1}{2}\phi^2$ when $\phi$ is small\cite{Bosch2015Analysis}. (\textbf{e}) dCoM with different defocus. (\textbf{f}) iCoM with different defocus. }
	\label{Fig CTF Virtual Image}
\end{figure}

The contrast of imaging modes depends on experimental parameters and can be explained using the contrast transfer functions (CTF). The 4D-Explorer software provides a real-time calculation of the CTF for different virtual imaging modes, which can guide users in selecting reconstruction parameters and interpreting the results. Figure \ref{Fig CTF Virtual Image} shows the analytical results of the CTF for various modes.

The calculations of CTFs of virtual imaging follows \cite{Bosch2015Analysis} and can be found in Supplementary Information \ref{Appendix Virtual Image}.  Figure \ref{Fig CTF Virtual Image}\textbf{a} displays the CTF for Axial BF, which is consistent with the one of the reverse-path TEM, as mentioned in the rotational offset correction of the calibration step. The transfer function of TEM is truncated at the spatial frequency corresponding to the radius of the bright-field diffraction disk, because we use the radius of the convergence aperture in STEM as the objective aperture radius in TEM. Figure \ref{Fig CTF Virtual Image}\textbf{b} shows the CTF for ABF, which exhibits an inverted contrast as defocus changes sign, indicating its sensitivity to imaging conditions. Figure \ref{Fig CTF Virtual Image}\textbf{c} shows the CTF for eABF, which is the difference between ABF and MaBF. eABF has an enhanced contrast, as ABF and MaBF have opposite contrasts at the same defocus, which can also be observed in Figure \ref{Fig Virtual Images}\textbf{b} and Figure \ref{Fig Virtual Images}\textbf{c}. Figure \ref{Fig CTF Virtual Image}\textbf{d} displays the bright field image using a virtual detector covering the full bright field disk (Full BF) and ADF. Note that Full BF and ADF do not have a contrast transfer under weak phase object approximation (WPOA), so their CTFs are calculated against $1-\cos\phi\approx\frac{1}{2}\phi^2$, where $\phi$ is the phase-change function of the sample. This also means that Full BF and ADF have lower contrast for light element samples. Unlike ABF, however, Full BF and ADF do not exhibit a contrast inversion with changing defocus sign, resulting in more stable imaging.

The calculation of the CTFs for the Center of Mass imaging mode is based on \cite{Lazic2016Phase} and can be found in Supplementary Information \ref{Appendix DPC}. Figure \ref{Fig CTF Virtual Image}\textbf{e} displays the CTF of dCoM with different defocus values. Due to the differential operations involved in the calculation of dCoM, the CTF curve of dCoM exhibits large absolute values. Figure \ref{Fig CTF Virtual Image}\textbf{f} shows the CTF of iCoM. Although the shape of the CTF curve of iCoM is similar to that of ADF, the CTF of iCoM is based on $\phi$ rather than $\phi^2$, making it better suited for imaging light-element samples than ADF. Compared to the CTF of ABF, the CTF of iCoM does not exhibit contrast reversal at different defocus values, resulting in better interpretability of iCoM images.

\subsection{Output Generation} 

The reconstruction results, as data matrices, are stored in the same HDF5 file along with the original 4D-STEM dataset, and can be exported in multiple data formats. Advanced analysis that is customized to a specific dataset can then be performed with other 4D-STEM softwares. For visualization, the image and vector field results can be rendered and viewed directly in 4D-Explorer through different viewers, and they can also be exported as high-quality plots to many image file formats.

\section{Conclusion} \label{Section Conclusion}

In this paper, we have introduced 4D-Explorer, an open-source software tool designed for 4D-STEM data processing and analysis. The software provides a complete visual workflow for data processing. For calibration, it integrates rotational offset correction and diffraction alignment. We have proposed a novel method for finding the rotational offset angle using the Axial BF image of a defocused 4D-STEM dataset. This approach has been experimentally validated and shown to be effective in achieving high-quality imaging results. For image reconstruction, 4D-Explorer includes virtual imaging and CoM that can reveal rich information about the sample. It also integrates real-time calculation and display of the transfer function during image reconstruction, which facilitates the selection of optimal reconstruction parameters and interpretation of reconstruction results. Overall, the software is user-friendly, cross-platform and extensible, making itself a valuable tool for researchers of various fields using 4D-STEM techniques. 

In future, there are several potential avenues for further development of 4D-Explorer software. One direction is the integration of machine learning algorithms to automate and optimize data processing workflows. This could include automated calibration, as well as more advanced image reconstruction techniques that leverage the power of deep learning\cite{Chang2023Deep}. Another area for future development is the incorporation of additional imaging modes. For example, in low-dose 4D-STEM experiment, using event-driven detectors\cite{Pelz2021Real, Jannis2022Event} to produce sparse datasets can effectively reduce the dataset size and acquiring duration, and thus the sparse datasets processing could be supported by 4D-Explorer. Additionally, combining 4D-STEM with other imaging techniques such as electron energy loss spectroscopy (EELS) could enable even more comprehensive characterization of materials\cite{Song2018Hollow}.

Users can visit \href{https://pengwang.org/4d-explorer/}{our website} for standalone executable applications, tutorials, documents and example datasets. The source codes are available in the \href{https://github.com/ManifoldsHu/FourDExplorer}{Github repository}. Users and developers are welcome to try our software and provide us with opinions, suggestions and contributions.

\section*{Acknowledgements}
The authors would like to acknowledge funding from the National Natural Science Foundation of China (11874199, 21991134, 21934007, 92056117), University of Warwick Research Development Fund (RDF) 2021-22 Science Development Award, Natural Science Foundation of Jiangsu Province (BK20210187) and China
Postdoctoral Science Fundation (No. 2022M711564). The authors thank Weina Fang for the preparation of the Gold Nanoparticle-modified DNA sample.

\bibliographystyle{unsrt} 
\bibliography{./ref.bib}

\begin{table}[htbp]
	\begin{tabular}{llll}
		\toprule 
		Sample	& \multicolumn{2}{l}{Gold Nanoparticle-modified DNA} & Monolayer \ce{MoS_2}\\
		\midrule
		Dataset Name & GP (infocus) & GP (underfocus) & \ce{MoS_2}\\
		Accelerate Voltage ($\SI{}{\kilo\volt}$) & 60 & 60 & 60 \\
		$\alpha$ ($\SI{}{\milli\radian}$) & 22.5 & 22.5 & 22.5 \\
		Scanning Step Size ($\SI{}{\nano\meter}$) & 1.80 & 1.80 & 0.029\\
		Defocus ($\SI{}{\micro\meter}$) & - & 2.11 & -\\
		Dwell Time ($\SI{}{\milli\second}$) & 1.0 & 1.0 & 1.0 \\
		Diffraction Image Size & $128\times 128$ &  $128\times 128$ & $128\times 128$\\
		Scanning Positions & $256\times 256$& $256\times 256$& $128\times 128$\\
		Preset Scanning Rotation ($\SI{}{\degree}$) & 30 & 30 & 0\\
		\bottomrule
	\end{tabular}
	\centering
	\caption{Parameters of the experiment datasets.}
	\label{Table Optical configuration}
\end{table}

\appendix 


\newcommand{\beginsupplement}{
    \setcounter{table}{0}
    \renewcommand{\thetable}{S\arabic{table}}
    \setcounter{figure}{0}
    \renewcommand{\thefigure}{S\arabic{figure}}
    \setcounter{equation}{0}
    \renewcommand{\theequation}{S\arabic{equation}}
    \setcounter{section}{0}
    \renewcommand{\thesection}{\arabic{section}}
    \setcounter{footnote}{0}
}

\begin{center}
	\textbf{\large Supplementary Information}
\end{center}
\begin{center}
	\textbf{\Large 4D-Explorer: a visual software for 4D-STEM data processing and image reconstruction}
\end{center}
\begin{center}
	{\large Hu Yiming et al.}
\end{center}


\beginsupplement{} 

\section{Basic Formulation of 4D-STEM Imaging}
Here we give a brief description of the 4D-STEM theory. A detailed introduction to 4D-STEM can be found in \cite{Ophus2019Four}. The wave function of the electron beam, after emitting from the source, is modified by the condenser lens and the aperture, noted as 
\begin{equation}
	\psi(\vec{u}) = C_{A}(\vec{u})E(\vec{u})e^{-i\chi(\vec{u})}
\end{equation}
where $\vec{u}=(u_x, u_y)$ denotes the location of the back focal plane (BFP) of the condenser lens, $E(\vec{u})$ is the envelope function, and $\chi(\vec{u})$ is the aberration function. $A$ is a subset of $\mathbb{R}^2$ and defined as 
\begin{equation}
	A = \left\{\vec{u}\in\mathbb{R}^2\left|\abs{u}\leq\frac{\alpha}{\lambda}\right.\right\}
\end{equation}
where $\alpha$ is the convergent semi-angle, and $\lambda$ is the wavelength of the beam. $C_A$ is the characteric function\footnote{Characteristic functions are defined as $ C_S(\vec{r})=\begin{cases}
	1,\quad \vec{r}\in S\\
	0,\quad \vec{r}\notin S
\end{cases} $ for any subset $S\subset\mathbb{R}^2$.} of $A$. Hence, 
\begin{equation}
	C_{A}(\vec{u})=\begin{cases}
		1,\quad &|\vec{u}| \leq\alpha/\lambda\\
		0,\quad &|\vec{u}| > \alpha/\lambda 
	\end{cases}
\end{equation}
We can also denote $u_{\text{BF}}=\alpha/\lambda$, which is the radius of the bright field disk on the detector plane.

The wave front propagating to the image plane acts as the probe transmitting the sample. According to the scalar diffraction theory, it can be written as $\psi_{\text{in}}(\vec{r}) = \FT{\psi(\vec{u})}(\vec{r})$, where $\mathcal{F}$ denotes the Fourier transform\footnote{The Fourier transform is defined as $ \FT{f(\vec{r})}(\vec{u})=\int_{\mathbb{R}^2}f(\vec{r})e^{-2\pi i \vec{u}\cdot\vec{r}}\dd[2]{\vec{r}}$, and the inverse Fourier transform is $ \iFT{g(\vec{u})}(\vec{r})=\int_{\mathbb{R}^2}g(\vec{u})e^{2\pi i\vec{u}\cdot \vec{r}}\dd[2]{\vec{u}$}.}. The interaction between the electron beam and the thin specimen can be abstracted to input wave multiplying a transmission function $t(\vec{r})$ around the scanning point $\vec{r}_p$, 
\begin{equation}\label{Eq wavefront_out}
	\psi_{\text{out}}(\vec{r}; \vec{r}_p)=\psi_{\text{in}}(\vec{r})t(\vec{r};\vec{r}_p) = \psi_{\text{in}}(\vec{r})e^{i\phi(\vec{r}+\vec{r}_p)}
\end{equation}
where $\phi(\vec{r})$ is the modulating phase-change function of the sample, which is irrelative to the scanning point. For non-magnetic thin samples, the phase change is proportional to the projected potential 
\begin{equation}
	\phi(\vec{r})=\sigma\int V(\vec{r}, z) \mathrm{d}z = \frac{2\pi m e\lambda}{h^2}\int V(\vec{r}, z) \mathrm{d}z
\end{equation}
where $\sigma$ is called interaction parameter, $e$ is charge, $m$ is relativistic mass, and $h$ is Planck's constant.

Then, following the scalar far-field diffraction theory again, the wave propagating to the detector plane is written as 
\begin{equation}
	\psi_D(\vec{u};\vec{r}_p) = \FT{\psi_{\text{out}}(\vec{r};\vec{r}_p)}(\vec{u};\vec{r}_p) = \FT{\psi_{\text{in}}(\vec{r})](\vec{u})*\mathcal{F}[t(\vec{r};\vec{r}_p)}(\vec{u};\vec{r}_p)
\end{equation}
where $*$ is the convolution operator\footnote{The convolution operator is defined as $\sbk{f(\vec{r}')*g(\vec{r}')}(\vec{r})=\int_{\mathbb{R}^2}f(\vec{r}')g(\vec{r}-\vec{r}')\dd[2]{\vec{r}'}$.}.

The intensity that can be recorded by the detector is
\begin{equation}\label{Eq wavefront_detector}
	I_{D}(\vec{u};\vec{r}_p)=\abs{\psi_{D}(\vec{u}; \vec{r}_p)}^2
\end{equation}
Sampling $I_{D}$ produces a 4D-STEM dataset $I_{D}(n_x, n_y; m_x, m_y)$, where $(n_x, n_y)$ is the coordinate of one diffraction image, and $(m_x, m_y)$ indicates the specific diffraction image that corresponds to the scanning point $\vec{r}_p$.

\section{Virtual Imaging}\label{Appendix Virtual Image}
Virtual images are produced by integrating $I_{\text{D}}$ over circular or annular regions on the diffraction plane for each scanning point $\vec{r}_p$. The analysis of the contrast of virtual images can be performed by the same process as conventional STEM, and has been introduced in \cite{Pennycook2011Scanning}. Here for integrity we provide primary analysis and note the main properties. Detailed analysis of virtual image contrast can be found in \cite{Bosch2015Analysis}. 

We start with subsets $D =\fbk{\left.\vec{u}\in\mathbb{R}^2\right|u_{\text{min}}\leq |\vec{u}|\leq u_{\text{max}}}$ that describe virtual detectors, and $C_D(\vec{u})$ as their corresponding characteristic functions. For BF detectors, $u_{\text{min}}=0$ and $u_{\text{max}}\leq u_{\text{BF}}$. For ABF detectors, $u_{\text{min}}>0$ and $u_{\text{max}}\leq u_{\text{BF}}$. For ADF detectors, $u_{\text{max}}>u_{\text{min}}>u_{\text{BF}}$. Then, the reconstructed image can be written as 
\begin{equation}\label{Eq VI_reconstruction}
	I_{\text{VI}}(\vec{r}_p)=\int_{\vec{u}\in S_{\text{D}}} I_{\text{D}}(\vec{u};\vec{r}_p)\dd[2]{\vec{u}} = \int_{\mathbb{R}^2} I_{\text{D}}(\vec{u}; \vec{r}_p)C_D(\vec{u})d^2\vec{u}
\end{equation}
The 4D-Explorer software uses the discrete version of the Equation (\ref{Eq VI_reconstruction}) to reconstruct virtual images. 

Our analysis of virtual imaging follows \cite{Bosch2015Analysis}. Substituting $I_D$ with Equation (\ref{Eq wavefront_detector}), we have that 
\begin{equation}\label{Eq Virtual Image Integration}
	I_{\text{VI}}(\vec{r}_p) = \int_{\mathbb{R}^2}\int_{\mathbb{R}^2}\int_{\mathbb{R}^2} C_D(\vec{u})e^{-2\pi i \vec{u}\cdot\rbk{\vec{r}' - \vec{r}}}\psi_{\text{in}}^*(\vec{r})\psi_{\text{in}}(\vec{r}')e^{i\phi(\vec{r}' + \vec{r}_p)} e^{-i\phi(\vec{r} + \vec{r}_p)} \dd[2]{\vec{r}} \dd[2]{\vec{r}'}\dd[2]{\vec{u}}
\end{equation}
Substituting with $K(\vec{r};\vec{r}_p) = e^{i\phi(\vec{r} + \vec{r}_p)} - 1$, we have that 
\begin{equation}
	I_{\text{VI}}(\vec{r}_p) = \iiint_{\mathbb{R}^6} C_D \abs{\psi_{\text{in}}}^2 e^{-2\pi i \vec{u}\cdot\rbk{\vec{r}' - \vec{r}}} \rbk{1 + K(\vec{r}';\vec{r}_p) + K^*(\vec{r};\vec{r}_p) + K(\vec{r}';\vec{r}_p)K^*(\vec{r};\vec{r}_p)} \dd[2]{\vec{r}} \dd[2]{\vec{r}'} \dd[2]{\vec{u}}
\end{equation}
Since then, we can write the $I_{\text{VI}}$ into four terms
\begin{equation}
	I_{\text{VI}}(\vec{r}_p) = I_0(\vec{r}_p) + I_{K}(\vec{r}_p) + I_{K^*}(\vec{r}_p) + I_{KK^*}(\vec{r}_p)
\end{equation}

Now we analyze the four terms respectively. The first term is that 
\begin{equation}
	I_0 = \int_{\mathbb{R}^2}C_D(\vec{u})\abs{\FT{\psi_{\text{in}}}}^2 \dd[2]\vec{u}
\end{equation}
which is a constant and does not contribute to the contrast.

The next two terms are that 
\begin{equation}
	I_{K}(\vec{r}_p) = \int_{\mathbb{R}^2} \psi_{\text{in}}(\vec{r}) K(\vec{r};\vec{r}_p) \FT{C_D(\vec{u})\iFT{\psi_{\text{in}}^*(\vec{r}')}(\vec{u})}(\vec{r})\dd[2]{\vec{r}}
\end{equation}
and 
\begin{equation}
	I_{K^*}(\vec{r}_p) = \int_{\mathbb{R}^2} \psi_{\text{in}}(\vec{r})^* K^*(\vec{r};\vec{r}_p) \iFT{C_D(\vec{u})\FT{\psi_{\text{in}}(\vec{r}')}(\vec{u})(\vec{r})}\dd[2]{\vec{r}}
\end{equation}
For convenience, let 
\begin{equation}\label{Eq Virtual Image B}
	B(\vec{r}) = \psi_{\text{in}}(\vec{r})\FT{C_D(\vec{u})\iFT{\psi_{\text{in}}^* (\vec{r}')}(\vec{u})}(\vec{r})
\end{equation}
and we have that 
\begin{equation}
    \begin{aligned}
    I_{K}(\vec{r}_p) + I_{K^*}(\vec{r}_p) &= \int_{\mathbb{R}^2} K(\vec{r};\vec{r}_p) B(\vec{r})\dd[2]{\vec{r}} + \int_{\mathbb{R}^2} K^*(\vec{r};\vec{r}_p) B^*(\vec{r}) \dd[2]{\vec{r}} \\
    &= \frac{1}{2}\int_{\mathbb{R}^2}\rbk{\rbk{K+K^*}\rbk{B+B^*}+\rbk{K-K^*}\rbk{B-B^*}}\dd[2]{\vec{r}}
    \end{aligned}
\end{equation}
Note that $(B+B^*)/2$ and $(B-B^*)/(2i)$ are real part and imaginary part of $B$ respectively, hence 
\begin{equation}
	I_{K} + I_{K^*} = -2\int_{\mathbb{R}^2} (1-\cos\phi(\vec{r}+\vec{r}_p))\Re(B(\vec{r}))\dd[2]{\vec{r}} - 2\int_{\mathbb{R}^2}\sin\phi(\vec{r} + \vec{r}_p)\Im(B(\vec{r}))\dd[2]\vec{r}
\end{equation}
We can use the cross-correlation\footnote{Cross-correlation is defined as that for any function $f(\vec{r})$ and $g(\vec{r})$, $[f(\vec{r})\star g(\vec{r})](\vec{r}') = \int_{\mathbb{R}^2} f^*(\vec{r})g(\vec{r}' + \vec{r})\dd[2]{\vec{r}}$} to define that 
\begin{equation}
	I_C(\vec{r}_p) = -2[\Re(B(\vec{r}))\star(1-\cos\phi(\vec{r}))](\vec{r}_p)
\end{equation}
and 
\begin{equation}
	I_S(\vec{r}_p) = -2[\Im(B(\vec{r}))\star\sin\phi(\vec{r})](\vec{r}_p)
\end{equation}
which are linear responses of $1-\cos\phi$ and $\sin\phi$ as the object respectively. Now using $I_C$ and $I_S$, the virtual imaging can be rewritten as
\begin{equation}
	I_{\text{VI}}(\vec{r}_p) = I_0 + I_C(\vec{r}_p) + I_S(\vec{r}_p) + I_{KK^*}(\vec{r}_p)
\end{equation}

The last term $I_{KK^*}$ is that 
\begin{equation}\label{Eq Virtual Image IKK Definition}
    I_{KK^*}(\vec{r}_p) = \int_{\mathbb{R}^2}\int_{\mathbb{R}^2} \psi_{\text{in}}^*(\vec{r})\psi_{\text{in}}(\vec{r}') K(\vec{r}';\vec{r}_p) K^*(\vec{r}; \vec{r}_p) \FT{C_D(\vec{u})}(\vec{r}' - \vec{r})\dd[2]{\vec{r}}\dd[2]{\vec{r}'}
\end{equation}
Substituting with $\vec{r}'' = \vec{r}' - \vec{r}$, the $I_{KK^*}$ can written in the form of auto-correlation
\begin{equation}
	I_{KK^*}(\vec{r}_p) = \int_{\mathbb{R}^2} \sbk{\psi_{\text{in}}(\vec{r})K(\vec{r};\vec{r}_p) \star \psi_{\text{in}}(\vec{r})K(\vec{r};\vec{r}_p)}(\vec{r}''; \vec{r}_p) \FT{C_D(\vec{u})}(\vec{r}'')\dd[2]\vec{r}''
\end{equation}

Further analysis requires additional approximation. According to \cite{Bosch2015Analysis}, the auto-correlation of $\psi_{\text{in}}K$ will reach its maximum at $\vec{r}''=\vec{0}$, but rapidly decays to $0$ outside a very small radius. Therefore, we use a new function to replace the auto-correlation, which remains the maximum value inside the small disk with radius $\rho$, while equals zero elsewhere. For amorphous carbon sample, $\rho = \SI{0.090}{\nano\meter}$, while for periodic system, $\rho = \SI{0.055}{\nano\meter}$. We use $C_{S_{\rho}}$ to denote the characteristic function of the small disk, and then the auto-correlation can be written as 
\begin{equation}
	\sbk{\psi_{\text{in}}(\vec{r})K(\vec{r};\vec{r}_p) \star \psi_{\text{in}}(\vec{r})K(\vec{r};\vec{r}_p)}(\vec{r}''; \vec{r}_p) \approx \abs{\psi_{\text{in}}(\vec{r})}^2 \abs{K(\vec{r}; \vec{r}_p)}^2 C_{S_\rho}(\vec{r}'')
\end{equation}
Hence, the $I_{KK^*}$ is that 
\begin{equation}
	I_{KK^*}(\vec{r}_p) \approx \sbk{\abs{\psi_{\text{in}}(\vec{r})}^2 \star 2(1-\cos\phi(\vec{r}))}(\vec{r}_p) \int_{\mathbb{R}^2} C_{S_\rho}(\vec{r}'')\FT{C_D(\vec{u})}(\vec{r}'') \dd[2]\vec{r}''
\end{equation}
which is also a linear response of $1-\cos\phi$. The integration 
\begin{equation}\label{Eq ADF Integration}
	G_D = \int_{\mathbb{R}^2} C_{S_\rho}(\vec{r}'')\FT{C_D(\vec{u})}(\vec{r}'') \dd[2]\vec{r}'' = J_0\rbk{\frac{2\pi}{\lambda} \beta_{\text{min}}\rho} - J_0\rbk{\frac{2\pi}{\lambda} \beta_{\text{max}}\rho}
\end{equation}
is a constant, in which $J_0$ is the Bessel function of the first kind and zero order. 

Since $I_{\text{VI}}$ can be regarded as linear responses of $\sin\phi$ and $1-\cos\phi$ as objects, we can write its contrast transfer function for each object. 

\paragraph{Detector Covering the Whole Plane} When the virtual detector covers all of the detector plane, i.e. $C_D = 1$ everywhere, the $B(\vec{r}) = \abs{\psi_{\text{in}}}^2$, and 
\begin{equation}
	I_{S}(\vec{r}_p) = 0,\quad I_{C}(\vec{r}_p) = -2 \sbk{\abs{\psi_{\text{in}}}^2 \star (1-\cos\phi(\vec{r}))}(\vec{r}_p)
\end{equation}
On the other hand, the last term of virtual imaging 
\begin{equation}
	I_{KK^*}(\vec{r}_p) = \int_{\mathbb{R}^2}\int_{\mathbb{R}^2} \psi_{\text{in}}^*(\vec{r})\psi_{\text{in}}(\vec{r}') K(\vec{r}';\vec{r}_p) K^*(\vec{r}; \vec{r}_p) \delta(\vec{r}' - \vec{r})\dd[2]{\vec{r}}\dd[2]{\vec{r}'}
\end{equation}
which is in fact 
\begin{equation}
	I_{KK^*} = 2 \sbk{\abs{\psi_{\text{in}}}^2 \star (1-\cos\phi(\vec{r}))}(\vec{r}_p)
\end{equation}
and hence vanishes along with $I_C$. Therefore, in this case there is no contrast at all
\begin{equation}
	I_{\text{Full Plane}}(\vec{r}_p) = I_0
\end{equation}
which is in our expectation because we have collected all of the electrons from the beam.

\paragraph{(Annular) Bright Field} When the virtual detector covers part of the bright field disk, i.e. $0\leq\beta_{\text{min}}<\beta_{\text{max}}<\alpha$, BF and ABF images can be generated. Specifically, for $\beta_{\text{min}}=0$ and $\beta_{\text{max}}=\alpha/2$, MaBF images are generated. Using weak phase object approximation (WPOA), we have that $\sin\phi \approx \phi$ and $1-\cos\phi \approx 1$, so only $I_S$ contributes to the contrast. The image of bright field and annular bright field can be written as 
\begin{equation}
	I_{\text{BF}}(\vec{r}_p) = I_0 - 2\sbk{ \Im\rbk{\psi_{\text{in}}(\vec{r})\FT{C_D(\vec{u}) \iFT{\psi_{\text{in}}^*(\vec{r}')}(\vec{u})}(\vec{r})} \star \sin\phi(\vec{r})}(\vec{r}_p)
\end{equation}
and the contrast transfer function is 
\begin{equation}\label{Eq CTF Bright Filed}
	T_{\text{BF}}(\vec{u}) = -2\mathcal{F}^*[\Im(\psi_{\text{in}} \mathcal{F}[C_D\mathcal{F}^{-1}[\psi_{\text{in}}^*]])](\vec{u})
\end{equation}

\paragraph{Annular Dark Field} When the virtual detector does not coincide with the bright field disk, i.e. $0 < \alpha \leq \beta_{\text{min}} < \beta_{\text{max}}$, ADF images can be generated. Since $A(\vec{k})C_D(\vec{u})=0$, we have that 
\begin{equation}
	C_D(\vec{u})\mathcal{F}^{-1}[\psi_{\text{in}}^*] = C_D(\vec{u}) A(\vec{u}) E(\vec{u})e^{i\chi(\vec{u})} = 0
\end{equation}
Hence, only $I_{KK^*}$ contributes to the contrast
\begin{equation}
	I_{\text{ADF}}(\vec{r}_p) = 2G_D\sbk{\abs{\psi_{\text{in}}(\vec{r})}^2\star (1-\cos\phi(\vec{r}))}(\vec{r}_p)
\end{equation}
where $G_D$ is calculated from Equation (\ref{Eq ADF Integration}). Here, if we use WPOA, we will not generate any contrast from ADF, which is the reason why it is hard for ADF to image light element samples. Thus we need to retain terms up to second order, $1-\cos\phi\approx\phi^2/2$ when $\phi$ is small. Using $1-\cos\phi$ as the sample, the contrast transfer function of ADF is 
\begin{equation}
	T_{\text{ADF}}(\vec{u}) = 2 G_D \mathcal{F}^*\left[|\psi_{\text{in}}|^2\right](\vec{u})
\end{equation} 

\paragraph{Full Bright Field} When the virtual detector covers the entire bright field diffraction disk, i.e. $\beta_{\text{min}} = 0$ and $\beta_{\text{max}} = \alpha$, the image is written as 
\begin{equation}
	I_{\text{Full BF}}(\vec{r}_p) = I_0 + \sbk{\rbk{-2\Re\rbk{\psi_{\text{in}}\FT{C_D\iFT{\psi_{\text{in}}^*}}} + 2G_D\abs{\psi_{\text{in}}}^2}\star(1-\cos\phi)}
\end{equation}
The contrast of Full BF is opposite to the ADF's contrast, because it can be regarded as the difference between the image obtained by a virtual detector covering only the plane outside the bright field diffraction disk, and $I_{\text{Full Plane}}$ which is a constant. Therefore, using $1-\cos\phi$ as the sample, its contrast transfer function is that 
\begin{equation}
	T_{\text{Full BF}}(\vec{r}_p) = 2\mathcal{F}^*\sbk{-\Re\rbk{\psi_{\text{in}}\FT{C_D\iFT{\psi_{\text{in}}^*}}} + G_D\abs{\psi_{\text{in}}}^2}
\end{equation}

\paragraph{Enhanced Annular Bright Field} According to \cite{Findlay2014Enhanced}, the enhanced annular bright field (eABF) is defined as the difference between the MaBF and ABF. Therefore, using WPOA, its image is that 
\begin{equation}
	I_{\text{eABF}}(\vec{r}_p) = I_{0} - 2\sbk{\Im\rbk{\psi_{\text{in}}\FT{\rbk{C_{D_{\text{ABF}}} - C_{D_{\text{MaBF}}}} \iFT{\psi_{\text{in}}^*}}} \star\sin\phi}(\vec{r}_p)
\end{equation}
and its contrast transfer function is that 
\begin{equation}
	T_{\text{eABF}}(\vec{u}) = -2\mathcal{F}^*\sbk{\Im\rbk{\psi_{\text{in}}\FT{\rbk{C_{D_{\text{ABF}}} - C_{D_{\text{MaBF}}}} \iFT{\psi_{\text{in}}^*}}}}(\vec{u})
\end{equation}


\section{Axial Bright Field}\label{Appendix Axial BF}
The depth of field of BF, ABF and ADF is usually narrow compared to CTEM images. However, Axial BF virtual detector produces images that have similar properties to CTEM, including a wide depth of field. This is from the principal of reciprocity \cite{Pogany1968Reciprocity, Bosch2015Analysis}. 

By substituting Dirac's delta function\footnote{Dirac delta function $\delta(\vec{r})$ is defined as that for any function $f(\vec{r})$,  $\int_{\mathbb{R}^2}f(\vec{r})\delta(\vec{r})d^2\vec{r} = f(\vec{0})$.} $\delta(\vec{u})$ to the virtual detector in Equation (\ref{Eq VI_reconstruction}), we have that 
\begin{equation}
	I_{\text{Axial BF}}(\vec{r}_p) = \int_{\mathbb{R}^2} I_{\text{D}}(\vec{u};\vec{r}_p)\delta(\vec{u})d^2\vec{u} = \abs{\int_{\mathbb{R}^2} \psi_{\text{in}}(\vec{r})e^{i\phi(\vec{r} + \vec{r}_p)}\dd[2]{\vec{r}}}^2
\end{equation}
which is exactly the same as the conventional flipped TEM image by substituting $\vec{r}_p$ with $-\vec{r}_p$
\begin{equation}
	I_{\text{Axial BF}}(-\vec{r}_p) = \abs{\sbk{e^{i\phi(\vec{r})} * \FT{\psi(\vec{u})}}(\vec{r}_p)}^2
\end{equation}
which has the same form as the image of HRTEM\cite{Williams2009Transmission}. By substituting $C_D(\vec{u})$ with $\delta(\vec{u})$ in Equation (\ref{Eq CTF Bright Filed})\cite{Bosch2015Analysis}, we have that 
\begin{equation}
    T_{\text{Axial BF}}(\vec{u}) = -2\mathcal{F}^*\sbk{\Im\rbk{\psi_{\text{in}} \FT{\delta(\vec{u}) \iFT{\psi_{\text{in}}^*}}}}(\vec{u}) = -2\mathcal{F}^*\sbk{\Im\rbk{\psi_{\text{in}}}}
\end{equation}
and hence 
\begin{equation}\label{Eq CTF AxialBF}
    T_{\text{Axial BF}}(\vec{u}) = 2 C_{A}(-\vec{u}) E(-\vec{u})\sin\rbk{\chi(-\vec{u})}
\end{equation}

In 4D-Explorer, we usually use a narrow bright field detector (for example, a single centered pixel of the diffraction image) to generate the axial BF images, and to approximate CTEM images.

\section{Center of Mass Imaging}\label{Appendix DPC}
There are two methods to produce differential phase contrast (DPC) \cite{Lazic2016Phase}. One is settling segmented virtual detectors and calculating differences of signals in the opposite orientations. This method is widely used, especially when pixelated electron detectors are not available. The other is to calculate the center of mass (CoM) of the bright field disk. It is also known as the first momentum imaging. In 4D-Explorer the default method to produce DPC vector fields is CoM, i.e. calculating center of mass of the bright field disk
\begin{equation}\label{Eq CoM_Definition}
	\vec{I}_{\text{CoM}}(\vec{r}_p) = \int_{\mathbb{R}^2}\vec{u} I_{\text{D}}(\vec{k};\vec{r}_p)\dd[2]{\vec{u}}
\end{equation}
which is a field distribution on the sample plane. Thus we can compute its divergence 
\begin{equation}\label{Eq dCoM_Definition}
	I_{\text{dCoM}}(\vec{r}_p) = \nabla_p\cdot\vec{I}_{\text{CoM}}(\vec{r}_p) = \frac{\partial I_{\text{CoM}x}(\vec{r}_p)}{\partial x_p} + \frac{\partial I_{\text{CoM}y}(\vec{r}_p)}{\partial y_p}
\end{equation}
as differentiated CoM (dCoM), and its potential $I_{\text{iCoM}}(\vec{r}_p)$ satisfying that
\begin{equation}\label{Eq iCoM_Definition}
	\vec{I}_{\text{CoM}}(\vec{r}_p) = \nabla_p I_{\text{iCoM}}(\vec{r}_p) = \left(\frac{\partial I_{\text{iCoM}}(\vec{r}_p)}{\partial x_p}, \frac{\partial I_{\text{iCoM}}(\vec{r}_p)}{\partial y_p}\right)
\end{equation}
as integrated CoM (iCoM), if the vector field is conserved. By Stokes theorem, the conservative field requires that the curl of the field (cCoM)
\begin{equation}\label{Eq cCoM_Definition}
	I_{\text{cCoM}}(\vec{r}_p)=\nabla_p\times\vec{I}_{\text{CoM}}(\vec{r}_p) = \frac{\partial I_{\text{CoM}y}(\vec{r}_p)}{\partial x_p} - \frac{\partial I_{\text{CoM}x}(\vec{r}_p)}{\partial y_p}
\end{equation}
equals zero everywhere.

Now we analyze the imaging of CoM. Following \cite{Lazic2016Phase}, we have that 
\begin{equation}
	I_{\text{CoM}x}(\vec{r}_p) = \int u_x\FT{\psi_{\text{out}}(\vec{r}); \vec{r}_p}(\vec{u}) \mathcal{F}^*\sbk{\psi_{\text{out}}(\vec{r}';\vec{r}_p)}(\vec{u})\dd[2]\vec{u}
\end{equation}
Then, using the derivative formula of Fourier transform, we have that 
\begin{equation}
	u_x\FT{f(\vec{r})} = \frac{1}{2\pi i}\FT{\pdv{f(\vec{r})}{x}}
\end{equation}
and hence 
\begin{equation}
	I_{\text{CoM}x}(\vec{r}_p) = \frac{1}{2\pi i}\int_{\mathbb{R}^2} \FT{\pdv{\psi_{\text{out}}(\vec{r};\vec{r}_p)}{x}}(\vec{u}) \FT{\psi_{\text{out}}^*(-\vec{r}'; \vec{r}_p)}(\vec{u}) \dd[2]{\vec{u}}
\end{equation}
Considering the fact that for any fucntion $f(\vec{r})$, its integration over the full plane 
\begin{equation}
	\int_{\mathbb{R}^2} f(\vec{r})\dd[2]{\vec{r}} = \FT{f(\vec{r})}(\vec{u} = 0)
\end{equation}
and using the convolution theorem, 
\begin{equation}
	I_{\text{CoM}x}(\vec{r}_p) = \frac{1}{2\pi i}\int_{\mathbb{R}^2}\pdv{\psi_{\text{out}}(\vec{r};\vec{r}_p)}{x}\psi_{\text{out}}^*(\vec{r};\vec{r}_p)\dd[2]\vec{r}
\end{equation}
Now substituting $\psi_{\text{out}}$ and $\pdv{\psi_{\text{out}}}{x}$ with $\psi_{\text{in}}$ and the transmission function of the sample, we have that 
\begin{equation}
	I_{\text{CoM}x}(\vec{r}_p) = \frac{1}{2\pi i}\int_{\mathbb{R}^2}\rbk{\pdv{\psi_{\text{in}}(\vec{r})}{x}\psi_{\text{in}}^*(\vec{r}) + i\psi_{\text{in}}(\vec{r})\psi_{\text{in}}^*(\vec{r})\pdv{\phi(\vec{r}+ \vec{r}_p)}{x}}\dd[2]\vec{r}
\end{equation}
which can be written in the form of cross-correlation 
\begin{equation}
	I_{\text{CoM}x}(\vec{r}_p) = \frac{1}{2\pi i}\int_{\mathbb{R}^2}\pdv{\psi_{\text{in}}(\vec{r})}{x} \psi_{\text{in}}^*(\vec{r})\dd[2]\vec{r} + \frac{1}{2\pi}\sbk{\abs{\psi_{\text{in}}(\vec{r})}^2 \star \pdv{\phi(\vec{r})}{x}}(\vec{r}_p)
\end{equation}
The first term is actually a constant and does not contribute to the contrast. Therefore, the relationship between CoM fields and the object is that
\begin{equation}\label{Eq CoM_contrast}
	\vec{I}_{\text{CoM}}(\vec{r}_p) = \frac{1}{2\pi}\sbk{\abs{\psi_{\text{in}}(\vec{r})}^2\star\nabla\phi(\vec{r})}(\vec{r}_p)
\end{equation}
which is a linear response to the gradient of the projected potential of the sample, i.e. the local projected electric field. The dCoM images can be written as 
\begin{equation}
	I_{\text{dCoM}}(\vec{r}_p) = \frac{1}{2\pi}\sbk{\abs{\psi_{\text{in}}(\vec{r})}^2\star \nabla^2\phi(\vec{r})}(\vec{r}_p)
\end{equation}
which is a linear response to the laplacian of the projected potential of the sample, i.e. the local projected charge density. Also, the iCoM image can be written as
\begin{equation}
	I_{\text{iCoM}}(\vec{r}_p) = \frac{1}{2\pi}\sbk{\abs{\psi_{\text{in}}(\vec{r})}^2\star \phi(\vec{r})}(\vec{r}_p)
\end{equation}
which is a linear response to the projected potential of the sample, i.e. the phase contrast. This is the reason why DPC images are sensitive to both light and heavy atoms simultaneously with good interpretability.

Next, we analyze the contrast transfer functions of CoM image modes. Using the cross-correlation theorem, we have that 
\begin{equation}
	T_{\text{iCoM}}(\vec{u}) = \frac{1}{2\pi}\mathcal{F}^* \sbk{\abs{\psi_{\text{in}}(\vec{r})}^2}(\vec{u})
\end{equation}
Based on the CTF of iCoM images and the differential and integral properties of Fourier transform, the CTF of CoM vector fields can be directly written as 
\begin{equation}
	\vec{T}_{\text{CoM}}(\vec{u}) = i\vec{u}\mathcal{F}^* \sbk{\abs{\psi_{\text{in}}(\vec{r})}^2}(\vec{u})
\end{equation}
Here, the $x$ and $y$ components of $\vec{T}_{\text{CoM}}(\vec{u})$ are the contrast transfer functions of the CoM vector field in the $x$ and $y$ directions, respectively. Similarly, the CTF of dCoM images (respect to the object $\phi$) can also be obtained
\begin{equation}
	T_{\text{dCoM}}(\vec{u}) = -2\pi \abs{u}^2 \mathcal{F}^* \sbk{\abs{\psi_{\text{in}}(\vec{r})}^2}(\vec{u})
\end{equation}

At last, we prove that the curl of CoM field is always zero. The first term in the definition of cCoM Equation (\ref{Eq cCoM_Definition}) is written as 
\begin{equation}
	\frac{\partial I_{\text{CoM}y}}{\partial x_p} = \frac{1}{2\pi}\int_{\mathbb{R}^2}\psi_{\text{in}}(\vec{r})\psi_{\text{in}}^*(\vec{r})\frac{\partial}{\partial x_p}\frac{\partial}{\partial y}\phi(x+x_p, y+y_p)d^2\vec{r}
\end{equation}
by substituting the contrast of CoM vector Equation (\ref{Eq CoM_contrast}). Assume that $\phi = \phi(u,v)$, where $u$ and $v$ denotes the first argument and the second argument of function $\phi$ respectively. Then, we have that
\begin{equation}\label{Eq cCoM_two_partial}
	\frac{\partial}{\partial x_p}\frac{\partial}{\partial y}\phi(x+x_p, y+y_p) = \frac{\partial}{\partial x_p}\left.\frac{\partial\phi(x+x_p, v)}{\partial v}\right|_{v=y+y_p}\frac{d(y+y_p)}{dy} = \left.\frac{\partial^2\phi(u,v)}{\partial u\partial v}\right|_{u=x+x_p, v=y+y_p}
\end{equation}

And due to the same reason of Equation (\ref{Eq cCoM_two_partial}), for the second term of Equation (\ref{Eq cCoM_Definition}) we have that 
\begin{equation}
	\frac{\partial}{\partial y_p}\frac{\partial}{\partial x}\phi(x+x_p, y+y_p) = \left.\frac{\partial^2\phi(u,v)}{\partial v\partial u}\right|_{u=x+x_p, v=y+y_p} =  \left.\frac{\partial^2\phi(u,v)}{\partial u\partial v}\right|_{u=x+x_p, v=y+y_p}
\end{equation}
Therefore, the two terms vanish for any $\vec{r}_p\in\mathbb{R}^2$, and  $I_{\text{cCoM}}(\vec{r}_p) = 0$.

\section{Rotational Offset Correction using CoM}\label{Appendix Rotaional Calibration}
Rotational offsets between the scanning coordinate and the detector coordinate will affect the reconstructed images if they are not carefully correctedd. Here, we provide the theory of correcting the rotational offset using properties of CoM vector fields. This method has the same physical origin as the J-Matrix method described in \cite{Ning2022Accurate}. 

Let $(\vec{e}_{px}, \vec{e}_{py})$ be the orthonormal basis of the scanning plane, and $(\vec{e}_{0x}, \vec{e}_{0y})$ be the orthonormal basis of the detector plane, as shown in Supplementary Figure \ref{Fig Rotational Offset Source}\textbf{a}. Then, a scanning position $\vec{r}_p$ can be written as $\vec{r}_p = x_p\vec{e}_x + y_p\vec{e}_y$, where $(x_p, y_p)$ is the coordinate of $\vec{r}_p$. Suppose there exists rotational offset with angle $\theta$ between the two planes such that 
\begin{equation}\label{Eq different_basis}
	\begin{cases}
		\vec{e}_{0x} = \vec{e}_{px}\cos\theta + \vec{e}_{py}\sin\theta\\
		\vec{e}_{0y} = -\vec{e}_{px}\sin\theta + \vec{e}_{py}\cos\theta
	\end{cases}
\end{equation}
At each scanning point $\vec{r}_p$ we receive a diffraction image, and we will calculate the center of mass of the images to reconstruct DPC images. If the rotational offset was not a major concern, we would use the following na\"{i}ve formula to calculate CoM vectors which is incorrect:
\begin{equation}\label{Eq naive_CoM}
	\vec{I}_{\text{nCoM}}(\vec{r}_p) = I_{\text{CoM}x}(\vec{r}_p)\vec{e}_{px} + I_{\text{CoM}y}(\vec{r}_p)\vec{e}_{py}
\end{equation}
That is because in computers the two directions of the scanning matrix and the detector matrix look the same. The subscript ``nCoM'' means the CoM vectors calculated by the na\"{i}ve method. However, as the camera rotated $\theta$ angle, the correct CoM vector at $\vec{r}_p$ is 
\begin{equation}\label{Eq correct_CoM}
	\vec{I}_{\text{CoM}}(\vec{r}_p) = I_{\text{CoM}x}(\vec{r}_p)\vec{e}_{0x} + I_{\text{CoM}y}(\vec{r}_p)\vec{e}_{0y}
\end{equation}
If we are not aware of $\theta$, no correct DPC reconstruction images can be generated, let alone other image modes such as ptychography or orientation mapping. 

Fortunately, DPC itself provides us with a method to correct the rotational offset. This method is based on the fact that CoM vector fields is linear to the projected electrostatic field of the sample, which should be conservative and irrotative, as stated in Equation (\ref{Eq CoM_contrast}). In this section we will prove that the rotational angle $\theta$ can be acquired by minimizing the sum of the squared curl of CoM vector field.

What we get through the detectors and CoM algorithms is the two components  $I_{\text{CoM}x}(\vec{r}_p)$ and $I_{\text{CoM}y}(\vec{r}_p)$ that should be a linear combination of detector's basis $\vec{e}_{0x}$ and $\vec{e}_{0y}$ as stated in Equation (\ref{Eq correct_CoM}). Therefore, the correct dCoM is 
\begin{equation}\label{Eq correct_dCoM}
	\begin{aligned}
		I_{\text{dCoM}}(\vec{r}_p) =&  \nabla_p\cdot\left(I_{\text{CoM}x}(\vec{r}_p)\vec{e}_{0x} + I_{\text{CoM}y}(\vec{r}_p)\vec{e}_{0y}\right)\\
		=& \nabla_p\cdot\left(I_{\text{CoM}x}(\vec{e}_{px}\cos\theta + \vec{e}_{py}\sin\theta) + I_{\text{CoM}y}(-\vec{e}_{px}\sin\theta + \vec{e}_{py}\cos\theta)\right)\\
		=& \left(\cos\theta\frac{\partial}{\partial x_p} + \sin\theta\frac{\partial}{\partial y_p}\right)I_{\text{CoMx}}(\vec{r}_p) + \left(-\sin\theta\frac{\partial}{\partial x_p} + \cos\theta \frac{\partial}{\partial y_p}\right)I_{\text{CoMy}}(\vec{r}_p)\\
		=& \left(\frac{\partial x_p}{\partial x_0}\frac{\partial}{\partial x_p} + \frac{\partial y_p}{\partial x_0}\frac{\partial}{\partial y_p}\right)I_{\text{CoM}x}(\vec{r}_p) + \left(\frac{\partial x_p }{\partial y_0}\frac{\partial}{\partial x_p} + \frac{\partial y_p}{\partial y_0}\frac{\partial}{\partial y_p}\right)I_{\text{CoM}y}(\vec{r}_p)\\
		=& \frac{\partial}{\partial x_0}I_{\text{CoM}x}(\vec{r}_p) + \frac{\partial}{\partial y_0}I_{\text{CoM}y}(\vec{r}_p)
	\end{aligned}
\end{equation}
This is expected because the divergence operator is not related to a specific basis, and that $I_{\text{CoM}x}$ and $I_{\text{CoM}y}$ are components of the detector basis. Due to the same reason, the correct curl of CoM is 
\begin{equation}
	I_{\text{cCoM}}(\vec{r}_p) = \frac{\partial}{\partial x_0}I_{\text{CoM}y}(\vec{r}_p) - \frac{\partial}{\partial y_0}I_{\text{CoM}x}(\vec{r}_p) = 0
\end{equation}
which has been shown in Section \ref{Appendix DPC}. 

Now we consider the wrong dCoM calculated using the na\"{i}ve method Equation (\ref{Eq naive_CoM}). We have that
\begin{equation}\label{Eq wrong_dCoM}
	\begin{aligned}
		I_{\text{ndCoM}}(\vec{r}_p) =& \nabla_p\cdot\left(I_{\text{CoM}x}(\vec{r}_p)\vec{e}_{px} + I_{\text{CoM}y}(\vec{r}_p)\vec{e}_{py}\right)\\
		=& \frac{\partial}{\partial x_p}I_{\text{CoM}x}(\vec{r}_p) + \frac{\partial}{\partial y_p}I_{\text{CoM}y}(\vec{r}_p)\\
		=& \frac{\partial x_0}{\partial x_p}\frac{\partial I_{\text{CoM}x}}{\partial x_0} + \frac{\partial y_0}{\partial x_p}\frac{\partial I_{\text{CoM}x}}{\partial y_0} + \frac{\partial x_0}{\partial y_p}\frac{\partial I_{\text{CoM}y}}{\partial x_0} + \frac{\partial y_0}{\partial y_p}\frac{\partial I_{\text{CoM}y}}{\partial y_0}\\
		=&\cos\theta\left(\frac{\partial I_{\text{CoM}x}}{\partial x_0} + \frac{\partial I_{\text{CoM}y}}{\partial y_0}\right) + \sin\theta\left(\frac{\partial I_{\text{CoM}y}}{\partial x_0} - \frac{\partial I_{\text{CoM}x}}{\partial y_0}\right)\\
		=& I_{\text{dCoM}}(\vec{r}_p)\cos\theta +  I_{\text{cCoM}}(\vec{r}_p)\sin\theta\\
		=& I_{\text{dCoM}}(\vec{r}_p)\cos\theta 
	\end{aligned}
\end{equation}
The result of the wrong dCoM shows that when $\theta = 0$, i.e. there is no rotational offset, the contrast of dCoM image reaches maximum. This property is key to our rotational offset correction algorithms. 

Meanwhile, we can also consider the wrong cCoM, which by similar process can be written as 
\begin{equation}\label{Eq wrong_cCoM}
	\begin{aligned}
		I_{\text{ncCoM}}(\vec{r}_p) =& \nabla_p\times\left(I_{\text{CoM}x}\vec{e}_{px} + I_{\text{CoM}y}(\vec{r}_p)\vec{e}_{py}\right)\\
		=& \frac{\partial}{\partial x_p}I_{\text{CoM}y}(\vec{r}_p) - \frac{\partial}{\partial y_p}I_{\text{CoM}x}(\vec{r}_p)\\
		=& \frac{\partial x_0}{\partial x_p}\frac{\partial I_{\text{CoM}y}}{\partial x_0} + \frac{\partial y_0}{\partial x_p}\frac{\partial I_{\text{CoM}y}}{\partial y_0} - \frac{\partial x_0}{\partial y_p}\frac{\partial I_{\text{CoM}x}}{\partial x_0} - \frac{\partial y_0}{\partial y_p}\frac{\partial I_{\text{CoM}x}}{\partial y_0}\\
		=& \cos\theta\left(\frac{\partial I_{\text{CoM}y}}{\partial x_0} - \frac{\partial I_{\text{CoM}x}}{\partial y_0}\right) - \sin\theta\left(\frac{\partial I_{\text{CoM}y}}{\partial y_0} + \frac{\partial I_{\text{CoM}x}}{\partial x_0}\right)\\
		=& I_{\text{cCoM}}(\vec{r}_p)\cos\theta - I_{\text{dCoM}}(\vec{r}_p)\sin\theta\\
		=& - I_{\text{dCoM}}(\vec{r}_p)\sin\theta
	\end{aligned}
\end{equation}
The result of the wrong cCoM shows that when $\theta = 0$, i.e. there is no rotational offset, cCoM does not have any contrast, which has been illustrated in Section \ref{Appendix DPC}. However, if there is rotational offset, the contrast of the cCoM images is proportional to dCoM, and reaches maximum when $\theta=\pi/2$. In this case, the reconstructed CoM vector fields becomes a non-divergent field with vortices around atoms.

The results from Equation (\ref{Eq wrong_dCoM}) and Equation (\ref{Eq wrong_cCoM}) provide us with a method to correct the rotational offset, where the angle $\theta$ that minimize the square sum of curl of CoM calculated from the na\"{i}ve approach:
\begin{equation}\label{Eq RotationalOffset_DPC}
	\theta = \arg\min_{\theta}\int_{\mathbb{R}^2}|I_{\text{ncCoM}}(\vec{r}_p;\theta)|^2 d^2\vec{r}_p
\end{equation}
In practice, we try to rotate every CoM vector by an angle, and calculate its corresponding value of the target function. The rotation angle is the one minimizing the target function. Then, we can rotate every diffraction images with that angle and perform reconstruction processes as usual. Note that there exist two solutions $\theta$ generated by Equation (\ref{Eq RotationalOffset_DPC}) differing $\pi$ from each other. Only one $\theta$ of them is physically valid, while the other $\theta'$ corresponds to the case where atoms have negative charge density. A simple way to distinguish between the two solutions is to compare reconstructed iCoM image with ADF images. Since ADF images are not sensitive to rotational offset and have the contrast with the same sign as the projected potential, the $\theta$ that leads to the same contrast is the correct one.

\bibliographystyle{unsrt} 
\bibliography{./ref.bib}


\begin{figure}
    \includegraphics[width=0.6\linewidth]{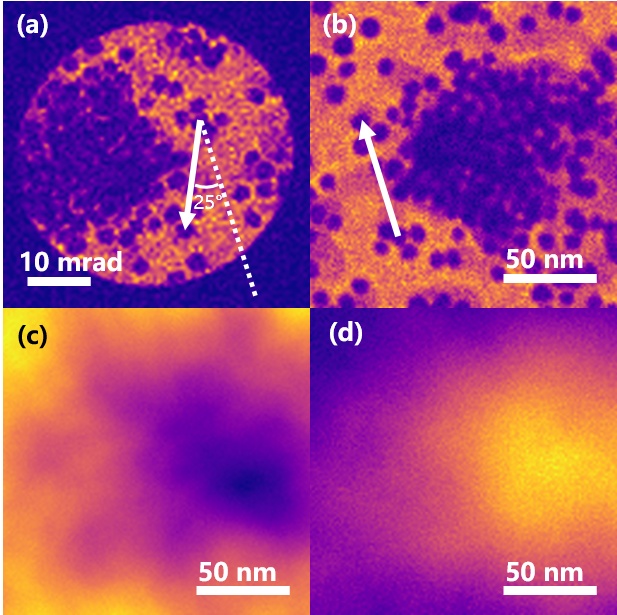}
    \centering
    \caption{The defocused 4D-STEM dataset of gold nanoparticles: diffraction pattern and virtual images. (\textbf{a}) A diffraction pattern of the 4D-STEM dataset. (\textbf{b}) Virtual axial bright field image. (\textbf{c}) Virtual bright field image. (\textbf{d}) Virtual dark field image. }
    \label{Fig Unblurred Axial BF}
\end{figure}

\begin{figure}
    \includegraphics[width=\linewidth]{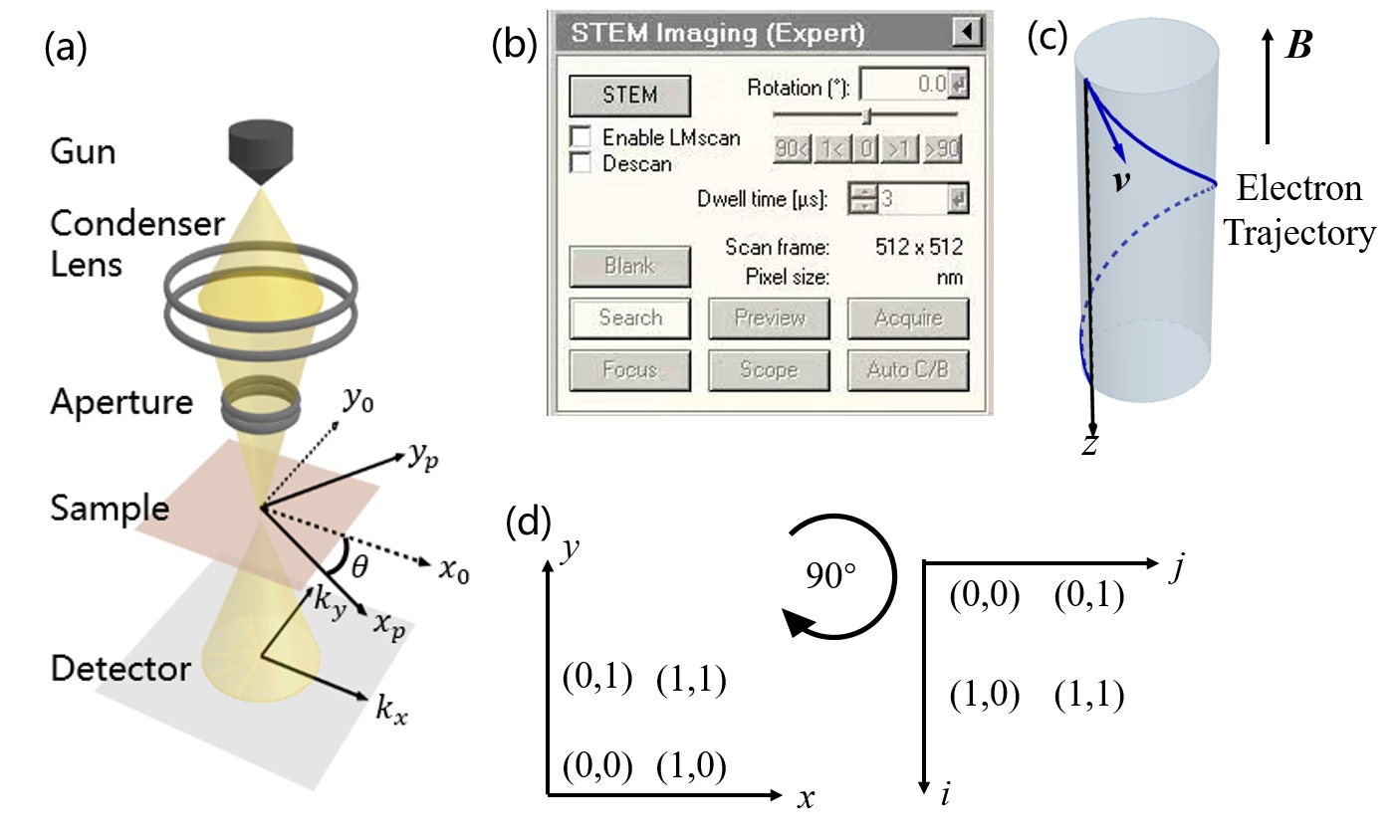}
    \centering
    \caption{The source of rotational offset. (\textbf{a}) The schema of the rotational offset, inducing the rotational angle $\theta$ between the scanning coordinate and the diffraction coordinate. (\textbf{b}) Operation panel of TEM Imaging \& Analysis (TIA). Users can manually set the scanning rotations during 4D-STEM experiments. (\textbf{c}) The helical trajectory of electrons caused by magnetic lenses. (\textbf{d}) Different choices of coordinate basis during computing. There exists a rotation of $\SI{90}{degree}$ between $x-y$ coordinates and $i-j$ indices.}
    \label{Fig Rotational Offset Source}
\end{figure}

\begin{figure}
    \includegraphics[width=\linewidth]{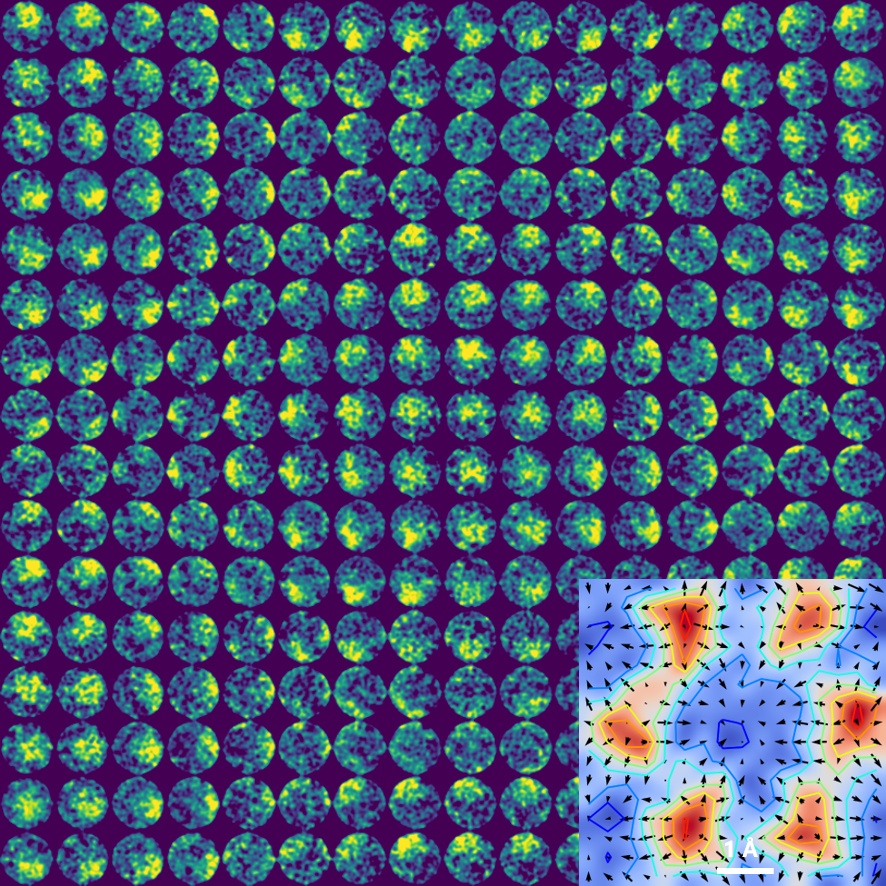}
    \centering
    \caption{Diffraction Patterns of the \ce{MoS_2} dataset. The center of mass of the patterns are shifted due to electric fields around the atom columns.}
\end{figure}

\begin{figure}
    \includegraphics[width=\linewidth]{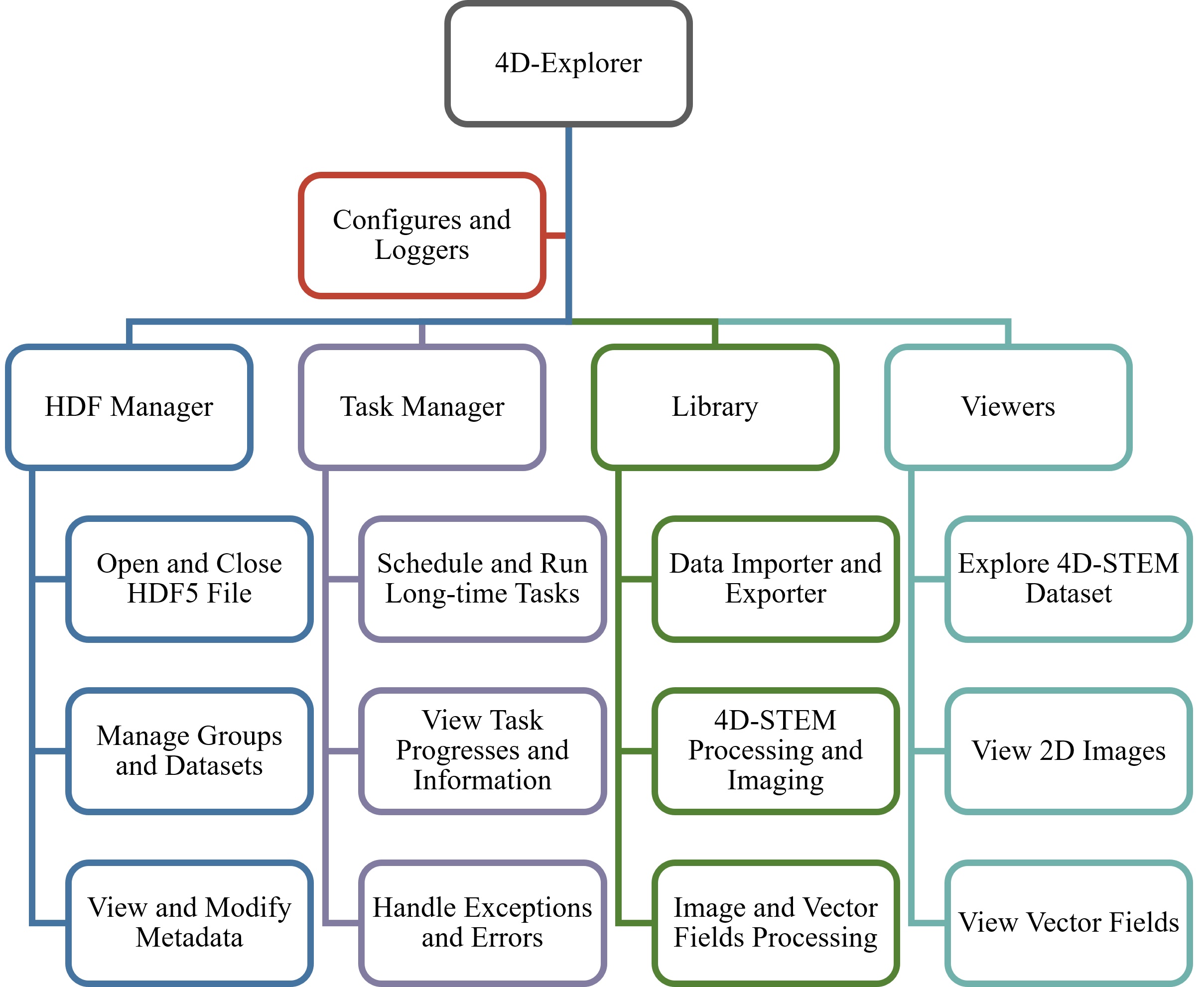}
    \centering
    \caption{The architecture and key components of 4D-Explorer and their key features. The hierarchy file format (HDF) manager handles data storage and their IO. The task manager keeps an executor with a thread pool for concurrent tasks, and will run these tasks automatically in the background without blocking the user interface. The library contains functions and algorithms, and encapsulate them into tasks. The viewers are actually a collection of dedicated displaying pages for different purposes. Apart from these components that directly related to 4D-STEM processing, other miscellaneous modules also support the functionality of the software, like the configuration manager and the loggers.}
    \label{Fig Architecture}
\end{figure}

\begin{figure}
    \includegraphics[width=\linewidth]{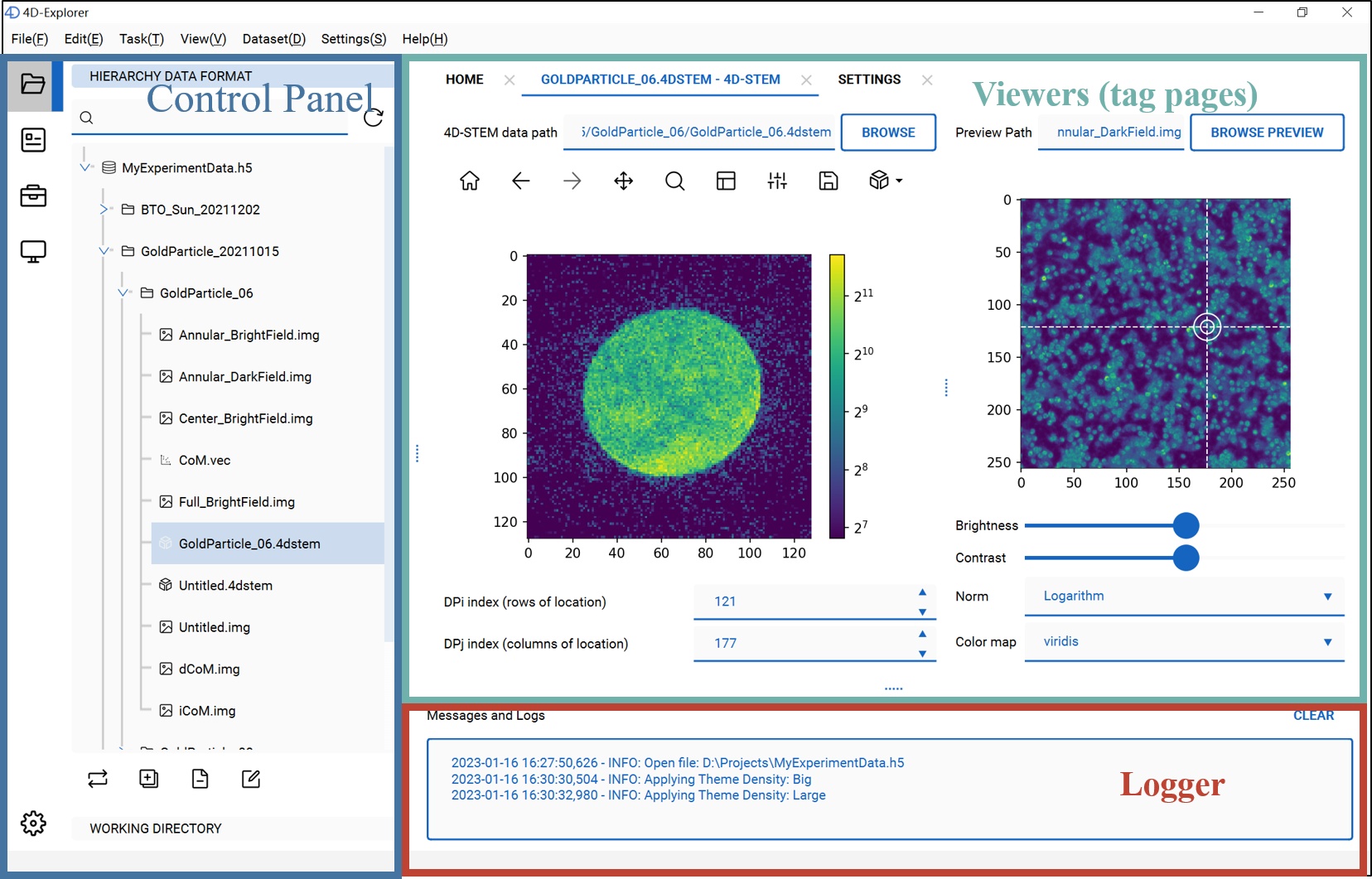}
    \centering
    \caption{The 4D-Explorer Main window. The control panel, the 4D-STEM viewer and the logger are displayed.}
    \label{Fig MainWindow Regions}
\end{figure}

\begin{figure}
    \includegraphics[width=\linewidth]{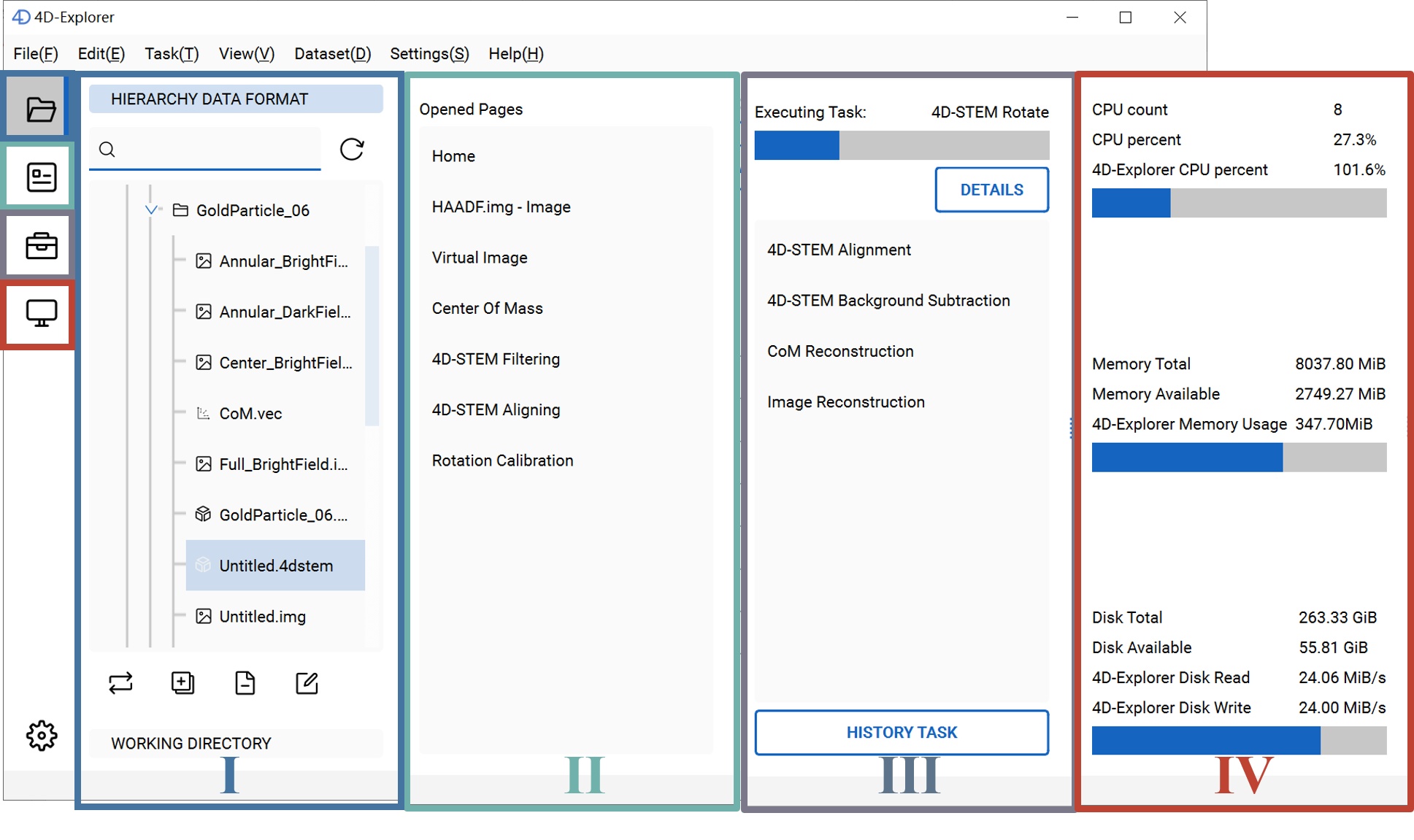}
    \centering
    \caption{Control panel. \textbf{I} File manager. Groups and datasets inside are displayed. \textbf{II} Currently opened viewers (tag pages). \textbf{III} Task Manager. Different compute tasks are executed in background. \textbf{IV} System information. }
    \label{Fig Control Panel}
\end{figure}

\begin{figure}
    \includegraphics[width=\linewidth]{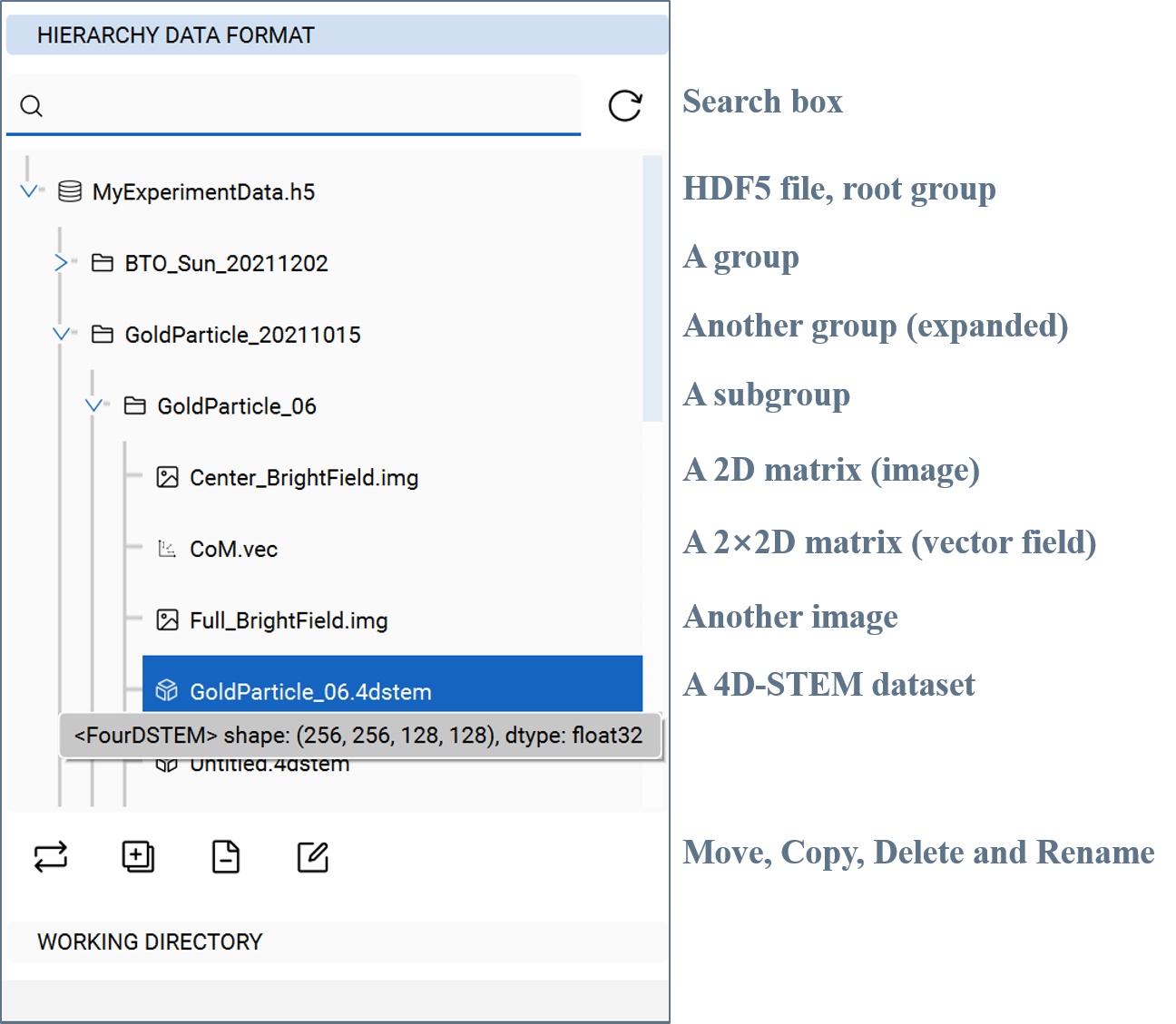}
    \centering
    \caption{File manager. The groups and datasets are organized as a tree in an HDF5 file. The HDF5 file here is \lstinline{MyExperimentData.h5}, and it is also the root group. Inside the file, there are several groups and subgroups. There are some datasets in a group \lstinline{GoldParticle_06}, which have different sizes and usage hinted by the extensions. For example, \lstinline{Center_BrightField.img} is a 2D matrix, can be viewed as an image, and has an extension \lstinline{.img}. Extensions include \lstinline{.vec}, \lstinline{.4dstem} and so on.}
    \label{Fig File Manager}
\end{figure}

\begin{figure}
    \includegraphics[width=\linewidth]{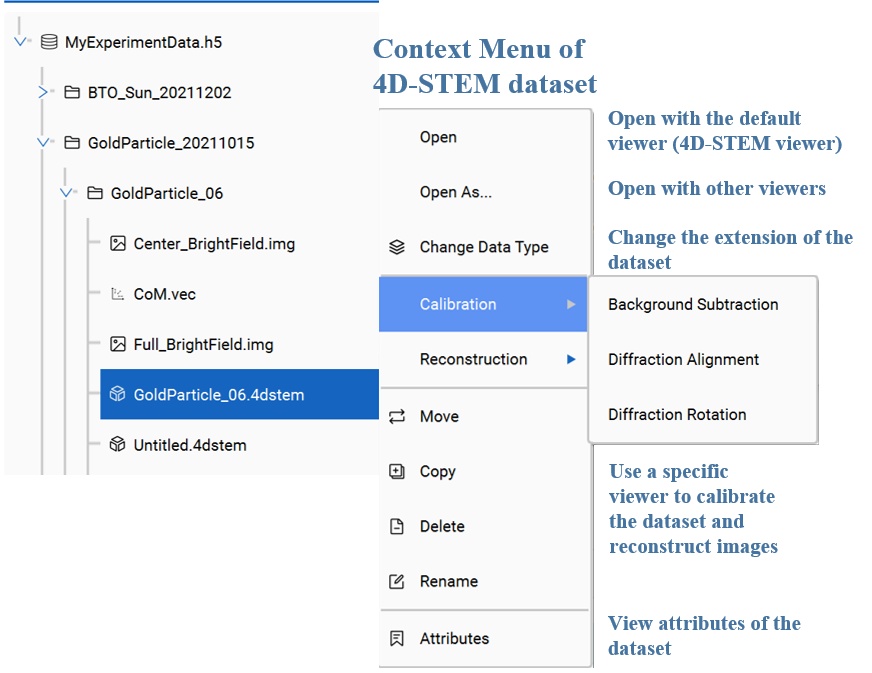}
    \centering
    \caption{A 4D-STEM dataset and its available actions. }
    \label{Fig Dataset}
\end{figure}

\begin{figure}
    \includegraphics[width=\linewidth]{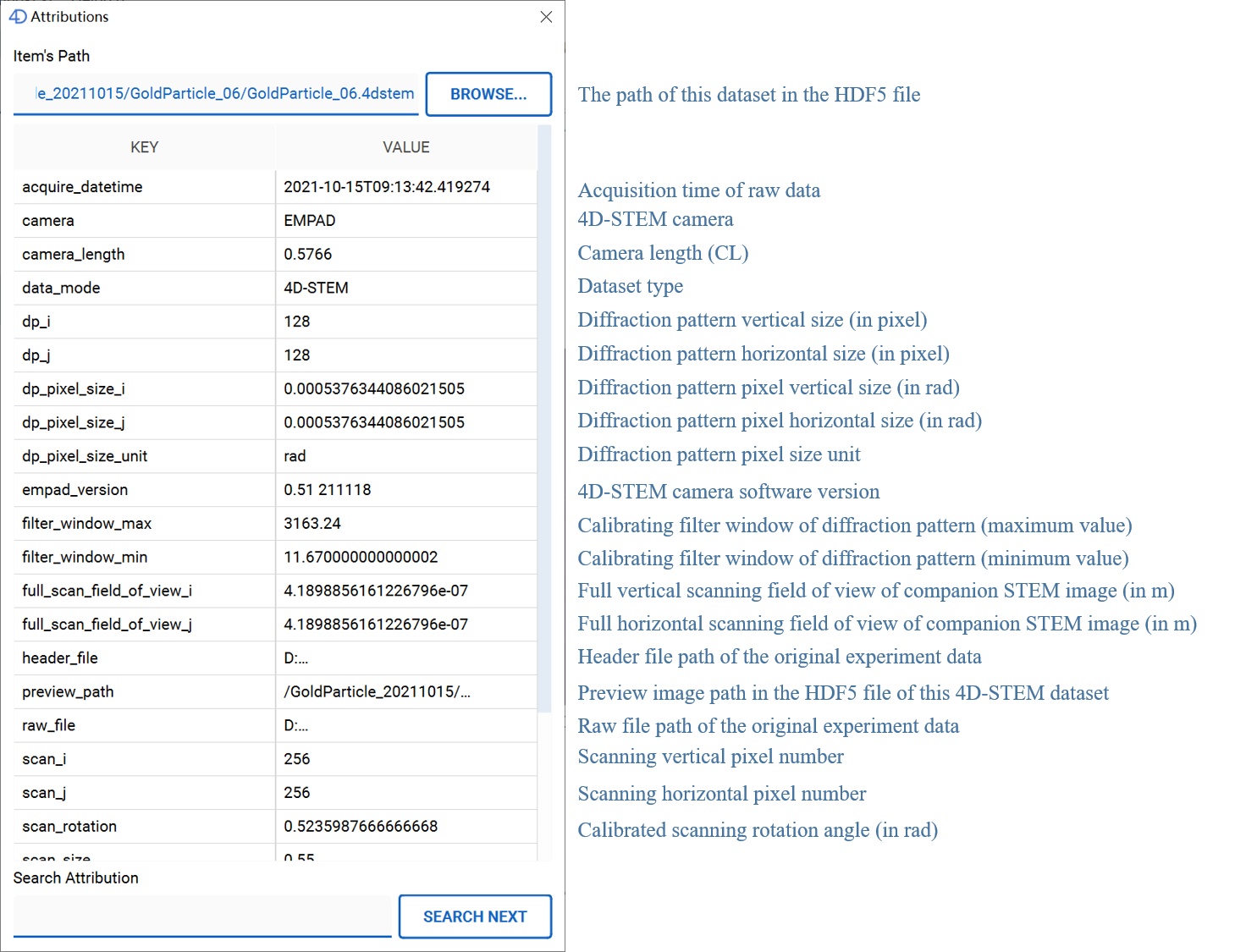}
    \centering
    \caption{Attributes of the 4D-STEM dataset in Figure \ref{Fig Dataset}.}
    \label{Fig Attributes}
\end{figure}

\begin{figure}
    \includegraphics[width=0.7\linewidth]{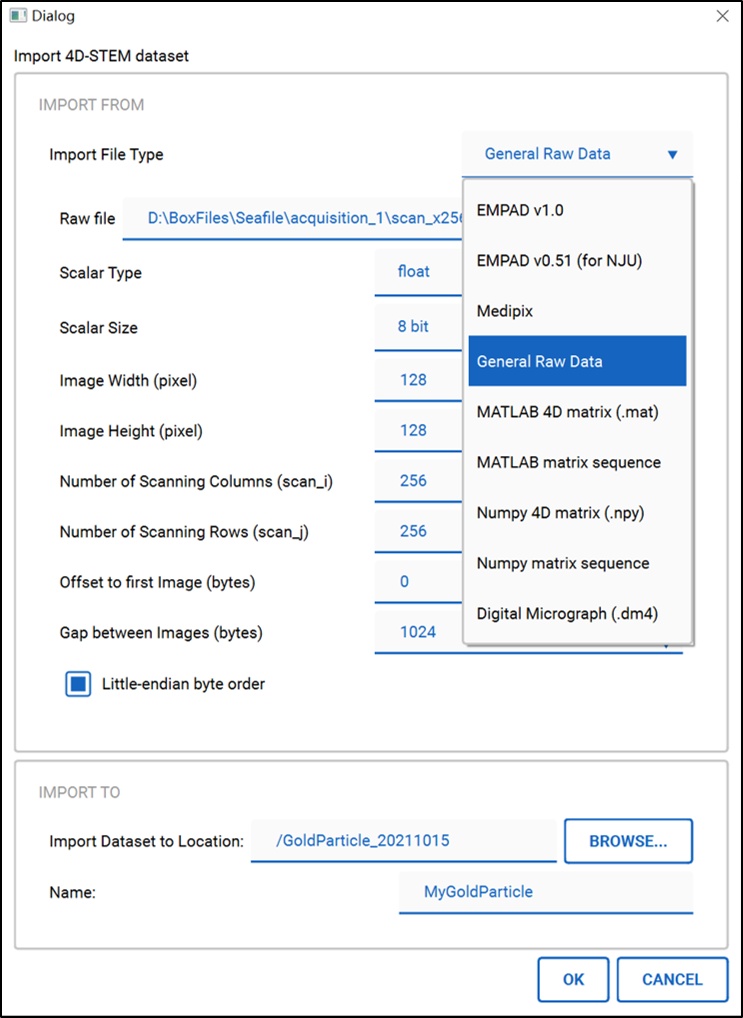}
    \centering
    \caption{Import 4D-STEM datasets. For general binary file (raw data), the reading arguments such as data type, height, width and offset can be customized. There are also predefined importer shown in the combobox. Dataset stored in the HDF5 file can also be exported as these kinds of files in similar processes.}
    \label{Fig Importer}
\end{figure}

\begin{figure}
    \includegraphics[width=\linewidth]{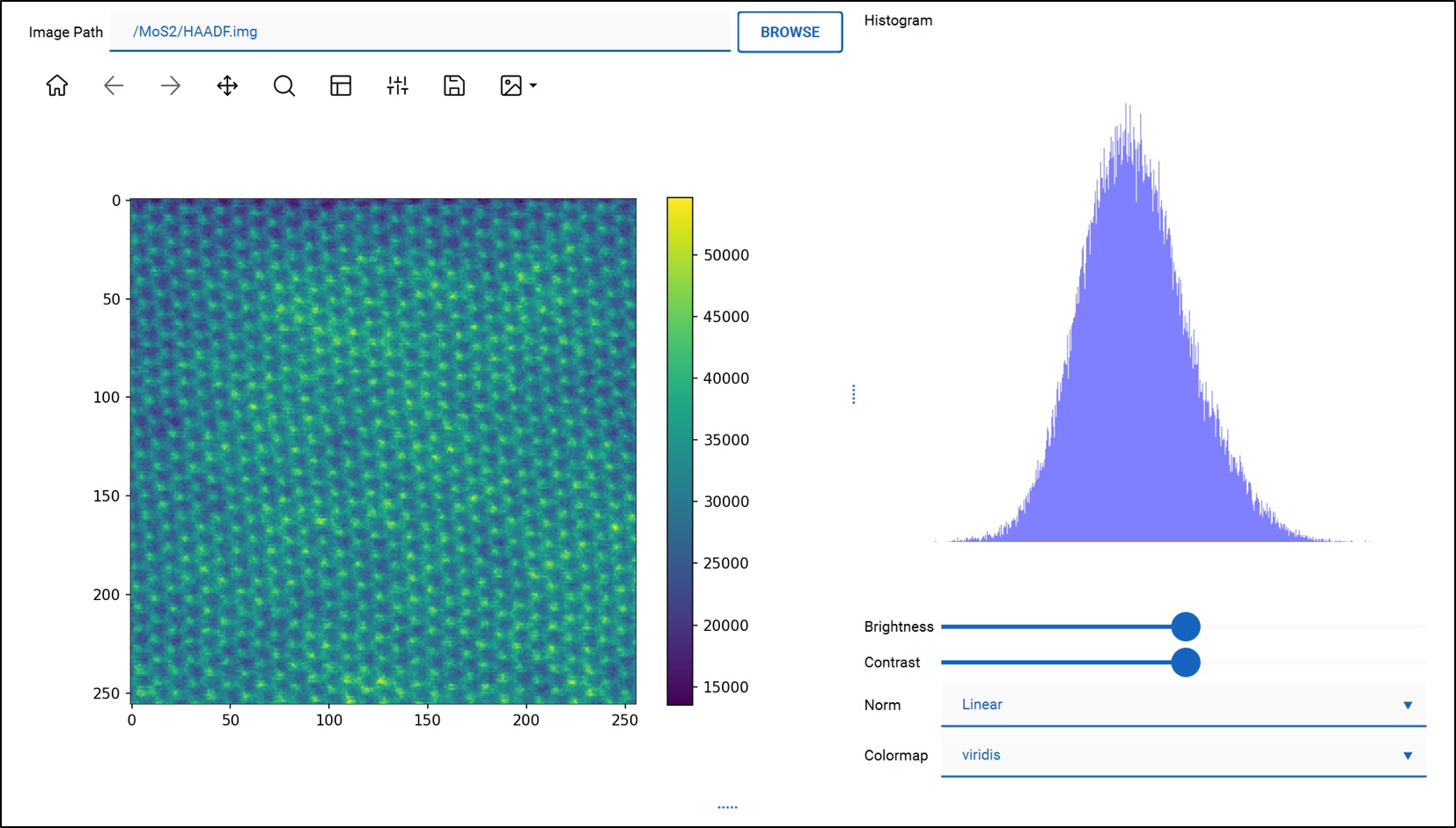}
    \centering
    \caption{The image viewer.}
    \label{Fig Image Viewer}
\end{figure}

\begin{figure}
    \includegraphics[width=\linewidth]{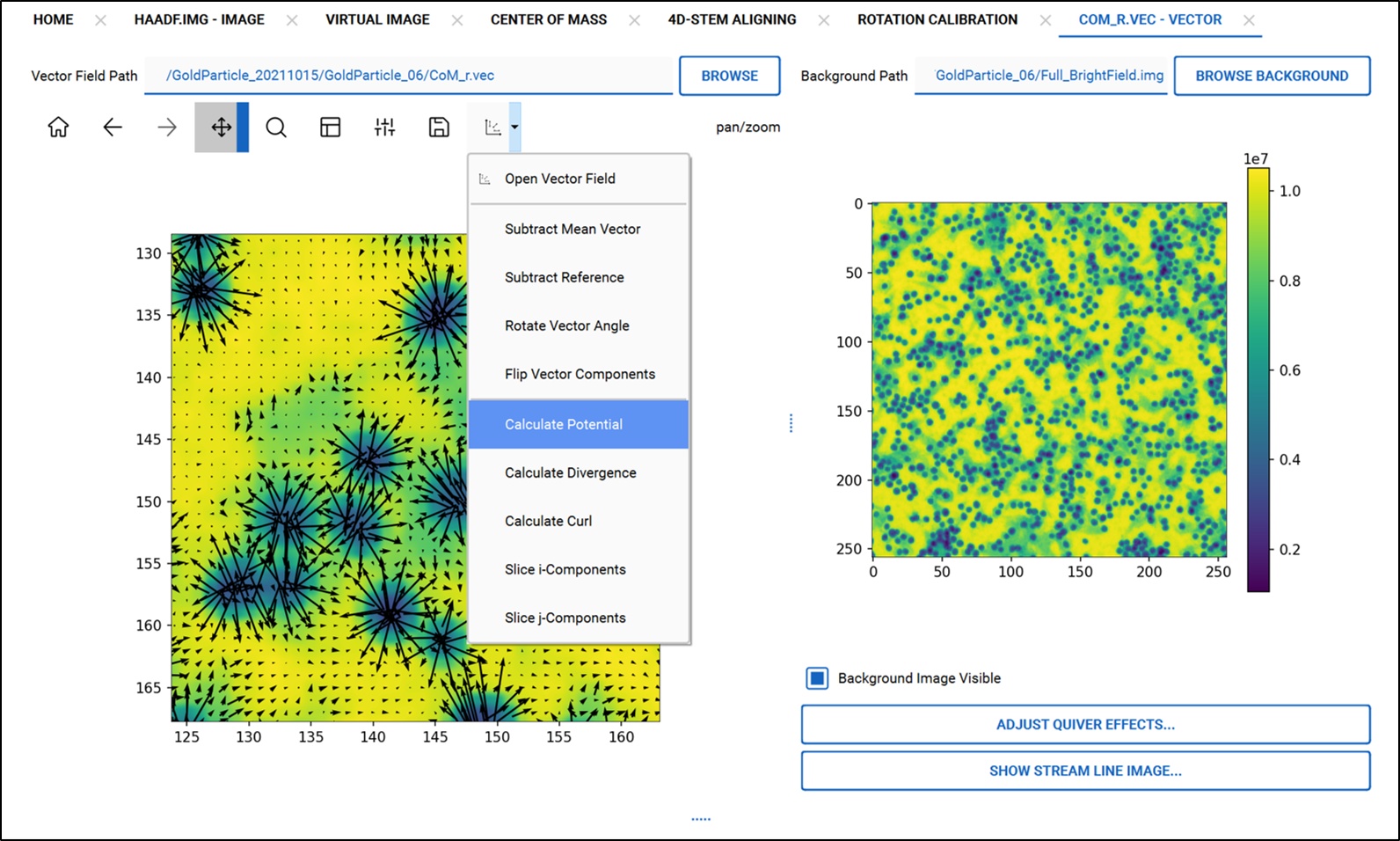}
    \centering
    \caption{The vector field viewer. The opened combobox here shows available actions to a vector field dataset. For a CoM distribution, we can compute its potential, divergence and curl to acquire iCoM, dCoM and cCoM respectively. }
    \label{Fig Vector Field Viewer}
\end{figure}

\begin{figure}
    \includegraphics[width=0.95\linewidth]{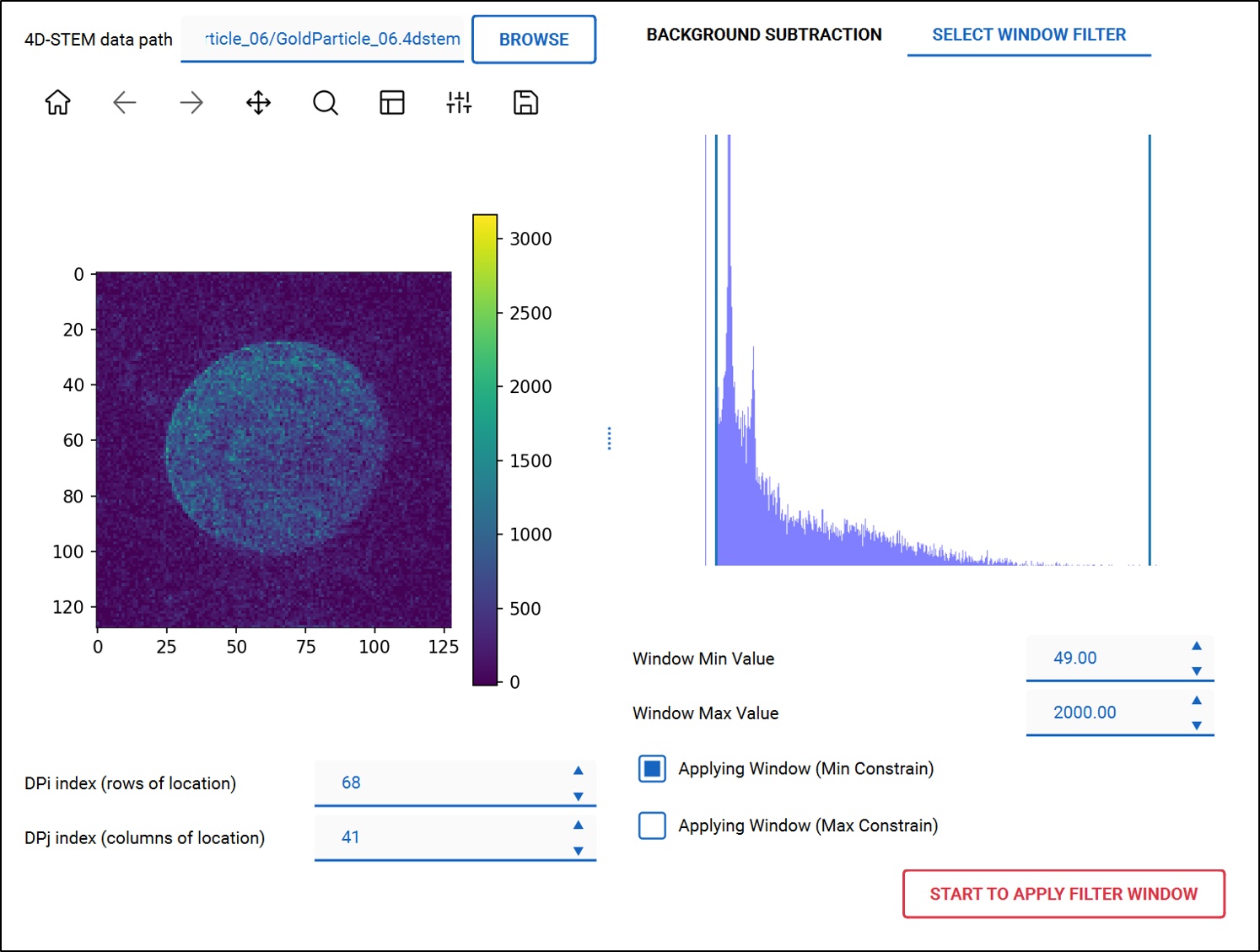}
    \centering
    \caption{The background subtraction viewer. The two blue vertical lines in the histogram are the two edges of the window filter.}
    \label{Fig Background Subtraction Viewer}
\end{figure}

\begin{figure}
    \includegraphics[width=\linewidth]{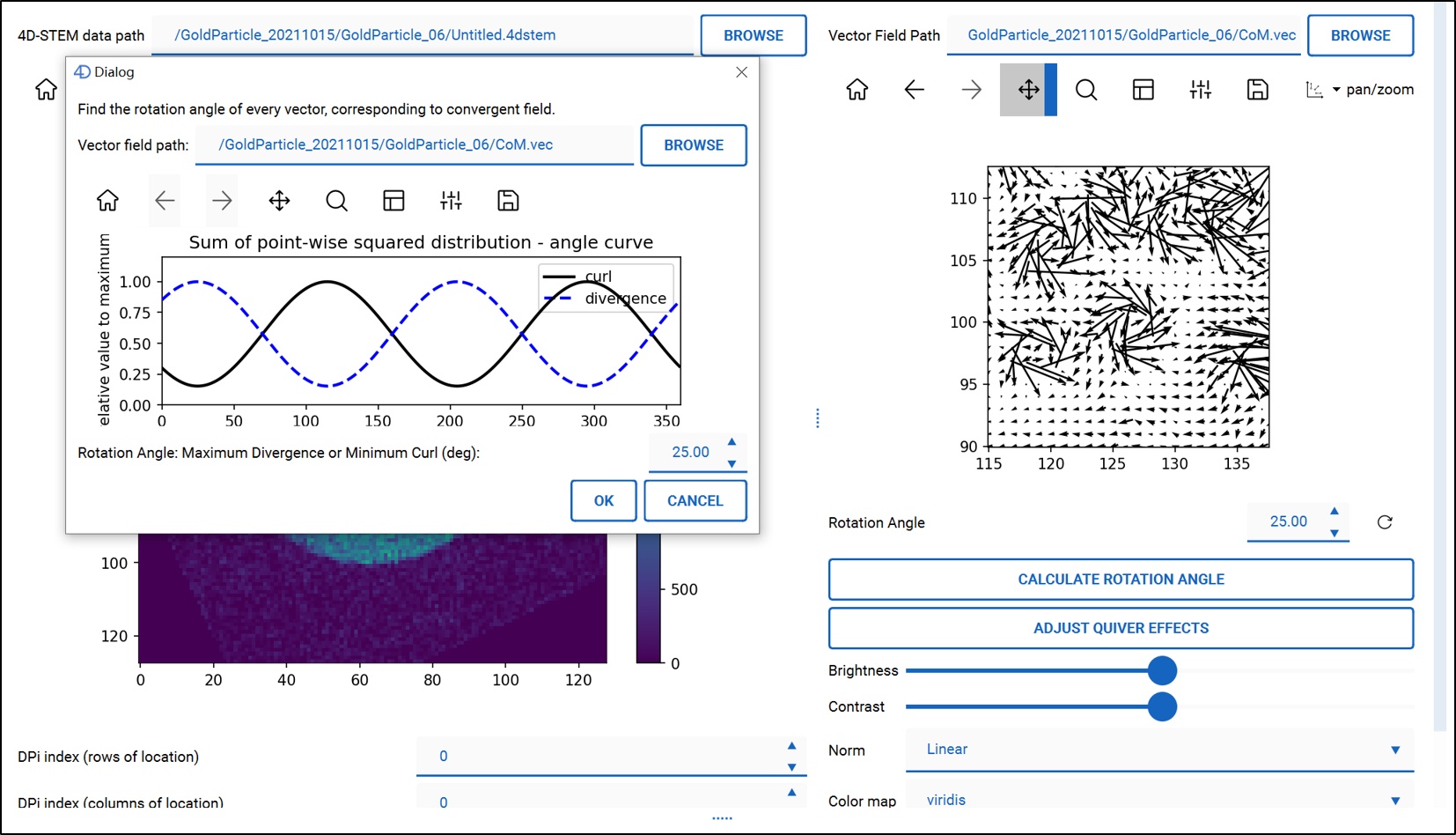}
    \centering
    \caption{The rotational offset correction viewer. We can solve the rotaional offset angle by minimizing the curl of CoM vector field. }
    \label{Fig Rotational Offset Viewer} 
\end{figure}

\begin{figure}
    \includegraphics[width=\linewidth]{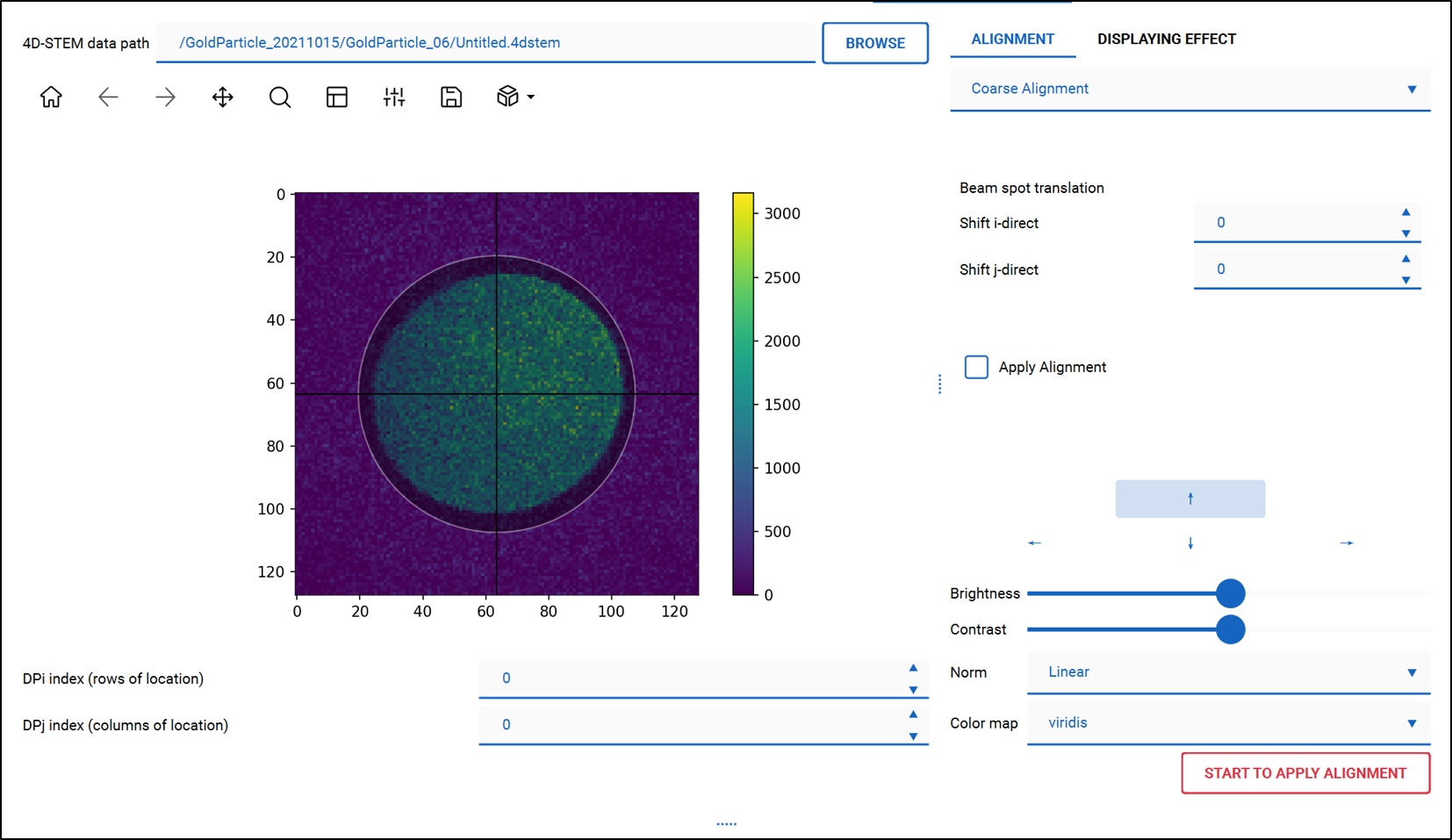}
    \centering
    \caption{The diffraction alignment viewer.}
    \label{Fig Diffraction Alignment Viewer}
\end{figure}

\begin{figure}
    \includegraphics[width=\linewidth]{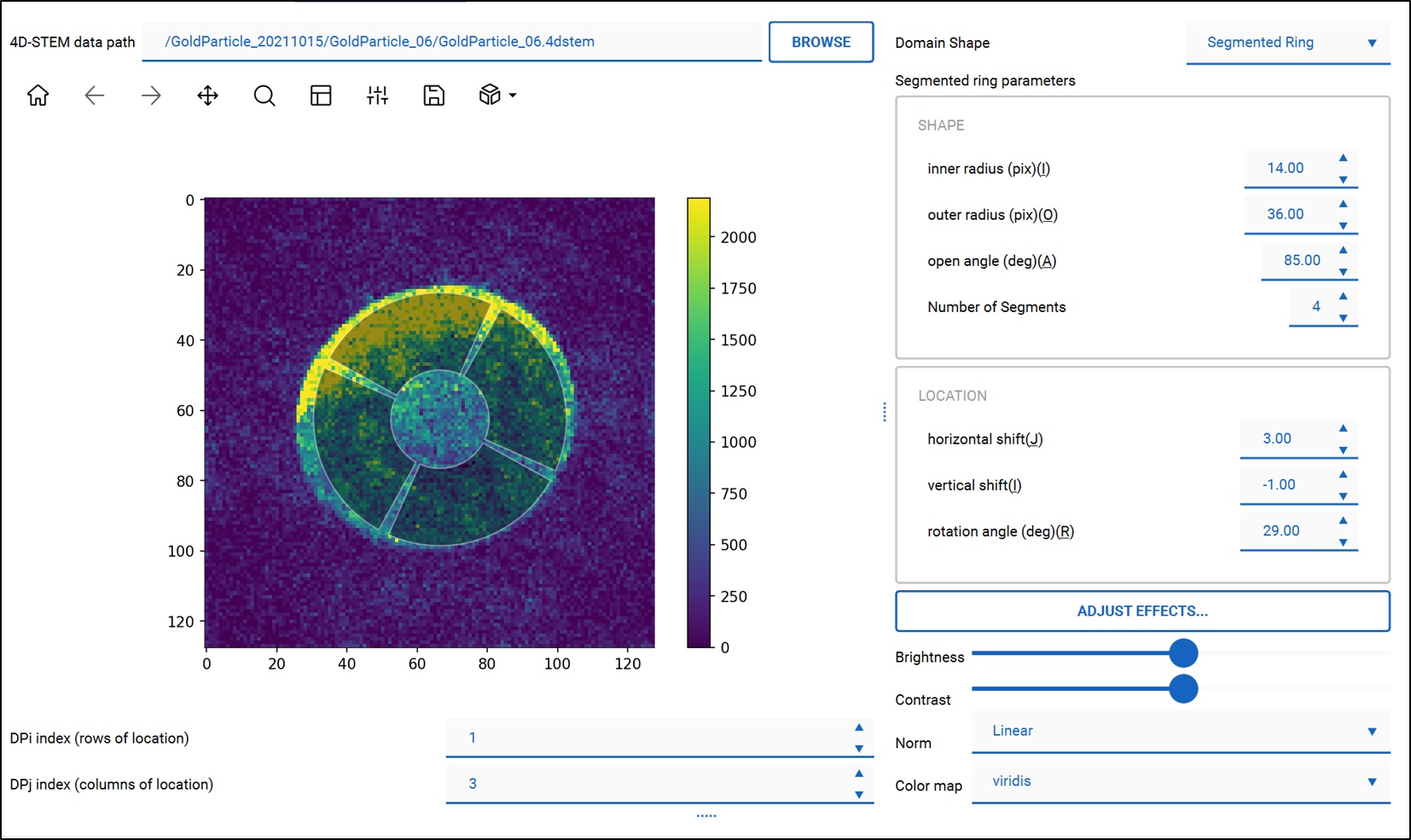}
    \centering
    \caption{The virtual image viewer. 4D-Explorer supports many virtual detector shapes for different purposes, including annulus, ellipse, rectangle and segmented wedges.}
    \label{Fig Virtual Image Viewer}
\end{figure}

\end{document}